\documentclass[aps,prd,twocolumn,showpacs,floatfix,
superscriptaddress,amsmath,preprintnumbers]{revtex4}

\usepackage{epsf}
\usepackage{amscd}              %
\usepackage{amsmath}            
\usepackage{bm}
\usepackage{latexsym}
\usepackage{graphicx}
\usepackage{comment}
\def\simge{
    \mathrel{\rlap{\raise 0.511ex
        \hbox{$>$}}{\lower 0.511ex \hbox{$\sim$}}}}
\def\simle{
    \mathrel{\rlap{\raise 0.511ex
        \hbox{$<$}}{\lower 0.511ex \hbox{$\sim$}}}}
\def\beqn{\begin{equation}}
\def\eeqn{\end{equation}}
\def\barr{\begin{eqnarray}}
\def\earr{\end{eqnarray}}
\def\bc{\begin{center}}
\def\ec{\end{center}}

\renewcommand{\l}{\left}
\renewcommand{\r}{\right}

\begin{document}
\preprint{DOE/ER/40762-373}
\title{Coulomb corrections in quasi-elastic scattering based on the eikonal expansion
for electron wave functions}

\author{J. A. Tjon}
\affiliation{Physics Department, University of Utrecht, 3508 TA Utrecht, The
Netherlands }
\affiliation{Department of Physics, University of Maryland, College Park,
MD 20742}

\author{S.~J.~Wallace}
 \affiliation{Department of Physics, University of Maryland, College Park,
              MD 20742}

\pacs{24.10.-i, 
      25.30.Fj, 
      25.30.Hm 
}

\date{\today}

\begin{abstract}
 An eikonal expansion is developed in order to provide systematic
corrections to the eikonal approximation through order $1/k^2$,
where $k$ is the wave number. The expansion is applied to wave
functions for the Klein-Gordon equation and for the Dirac
equation with a Coulomb potential. Convergence is rapid at
energies above about 250 MeV. Analytical results for the eikonal
wave functions are obtained for a simple analytical form of the
Coulomb potential of a nucleus. They are used to investigate
distorted-wave matrix elements for quasi-elastic electron
scattering from a nucleus. Focusing factors are shown to arise
from the corrections to the eikonal approximation. A precise form
of the effective-momentum approximation is developed by use of a
momentum shift that depends on the electron's energy loss.
\end{abstract}

\pacs{}

\maketitle


\section{Introduction}
\label{sec:intro}


Quasielastic scattering of electrons by nuclei can provide
important information about the response of a nucleus to a weakly
interacting probe.  Experiments have been performed at the MIT
Bates Laboratory ~\cite{Altemus83, Deady83, Hotta84, Deady86,
Blatchley86, Dytman86, Dow88, Yates93, Williamson97}, at the
Saclay Laboratory~\cite{Barreau83, Meziani84, Meziani85,
Marchand85, Zghiche94,Gueye99} and at SLAC~\cite{Baran88, Chen91,
Mezianai92} in order to explore this reaction. A recent review
of quasi-elastic scattering provides a guide to the experimental
and theoretical results~\cite{Benhar06}.  One important theoretical issue
concerns the corrections that arise from the Coulomb potential of
the nucleus.  Another issue is the accuracy with which
longitudinal (L) and transverse (T) response functions can be
determined when Coulomb corrections are present.

In order to include the Coulomb distortion of electron waves, it is
necessary to solve the Dirac equation for scattering of an
electron of mass $m$ and energy $E$ from a nucleus, i.e.,
\begin{equation}
\Big( E - \alpha \cdot {\bf p} - V(r) - \beta m\Big) \psi = 0,
\end{equation}
where $V(r)$ is the Coulomb potential due to the nuclear charge
distribution.   Exact solutions for the Dirac-Coulomb wave
functions may be obtained as a sum over partial
waves.~\cite{Jin9294,Udias93,Udias95} However, the
 sum involves delicate numerical computations.
 As the electron energy increases, the
partial-wave expansions converge more slowly in spite of the fact
that the Coulomb corrections become smaller.   A simpler approach
is to use the eikonal approximation for the distortions because
the approximation gets better as the energy increases and it can
provide insight into the parameters that control the Coulomb
corrections.

For waves moving along the z-direction, the eikonal approximation
provides an approximate
 solution,
 \begin{equation}
 \psi \approx e^{i k z} e^{i \chi},
 \end{equation}
 where $k=\sqrt{E^2-m^2}$ is the momentum
 and $\chi$ is a simple function of the potential.
 It is important to include focusing factors
such as those derived by Yennie, Boos and
Ravenhall~\cite{Yennie65, Lenz71,Rosenfelder80} based upon the
WKB approximation to the distorted Coulomb waves. Some particulary
transparent results have been obtained using the eikonal
approximation to derive an effective-momentum approximation
(ema)~\cite{Traini01,Traini88} that produces results very similar
to plane-wave results. However, with the combination of the
eikonal approximation, WKB focusing factors and the
effective-momentum approximation, the approach lacks a systematic
basis and accuracy is uncertain. Approximations to the
partial-wave analysis, such as the analysis of Ref~\cite{Kim96},
that attempt to improve the effective-momentum approximation also
have uncertain accuracy.  Significant disagreements in the
determination of nuclear response
functions from experimental data~\cite{Morgenstern01,Benhar06} have
arisen at
least in part owing to the use of different theoretical methods to
remove the Coulomb corrections. 

 In order to address the issue of Coulomb corrections, we develop
corrections to the eikonal approximation based upon a systematic
expansion about the high-energy limit. This eikonal expansion is
shown to be rapidly convergent for typical energies and targets
used in quasi-elastic scattering. For a simple analytical Coulomb
potential, analytical forms are developed for the eikonal
corrections through second-order in $1/k$, where $k$ is the
electron's wave number.

The eikonal approximation has a long history~\cite{Moliere47}.
Glauber used the eikonal approximation to develop a simple and
insightful form of multiple scattering theory that is called
Glauber theory~\cite{Glauber59}. Czyz and Gottfreid~\cite{Czyz63}
used the eikonal approximation to analyze electron scattering but
that work did not include focusing factors. Work by Giusti {\it et
al.} also is based on the eikonal
approximation~\cite{Giusti87,Giusti88} and some recents works
have combined the eikonal approximation with semi-classical
focusing factors in order to assess Coulomb corrections in
quasi-elastic scattering.~\cite{Aste04a,Aste04b}. We show that the
focusing factors arise from systematic corrections to the eikonal
approximation.

Corrections to the eikonal approximation also have a long
history.  Work by Saxon and Schiff~\cite{Saxon57} showed how to
correct the approximation to leading order in $1/k$. A systematic
expansion for the scattering t-matrix was developed by Sugar and
Blankenbecler~\cite{Sugar69}.  Systematic corrections to the
Glauber approximation were developed by Wallace~\cite{Wallace73}
and extended to the Dirac scattering amplitude
in~\cite{Wallace84}. However, a systematic expansion for wave
functions has not been developed prior to this work.

   In Section~\ref{sec:eikonal_KG}, we develop the eikonal expansion for Klein-Gordon
wave functions because that is the simplest and most transparent
case. In Section~\ref{sec:eikonal_Dirac},
we develop the eikonal expansion for the
Dirac wave function.  Section~\ref{sec:upper-components}
focuses on $u({\bf r})$, which is a Pauli
spinor containing the two upper components of the Dirac wave
function. Because there is a spin-orbit interaction, additional
spin-dependent terms arise in the eikonal expansion. Because the analysis
is technically more complicated, details are
given in an appendix. In Section~\ref{sec:lower-components}
 we show that the Pauli
spinor $\ell({\bf r})$ that contains the two lower components of
the Dirac wave function takes a very simple form in the limit of
vanishing electron mass, i.e., $\ell({\bf r}) = 2 \lambda u({\bf
r})$, where $\lambda = \pm {1\over 2}$ is the helicity. This
follows because of the structure of the Dirac equation in the
limit of vanishing electron mass. Conservation of the electron's
helicity holds to high accuracy at electron energies of interest.
Section~\ref{sec:convergence} discusses convergence of the eikonal expansion and
analytical results for the eikonal phases, which are given in an appendix.
Section~\ref{sec:focusing-factors} discusses the different focusing factors that
arise for Klein-Gordon and Dirac waves.
In Section~\ref{sec:currentME}, we consider helicity matrix elements of the electron current
based on the Dirac distorted waves.   We
show that the spin-orbit distortions arising from the Coulomb
interaction can alter the longitudinal and transverse helicity
matrix elements
in somewhat different ways, which can affect the L/T
separation. Specializing to the longitudinal response,
section~\ref{sec:quasi-elastic} discusses quasi-elastic scattering by use of a
simple model of the nuclear response.  Section~\ref{sec:ema} revisits the
effective-momentum approximation ($ema$) and Section~\ref{sec:numerical}
presents results of calculations of the longitudinal response function.
The analytical eikonal wave functions with
systematic corrections included are used to describe the
Coulomb corrections.
We show that the effective-momentum
approximation can be made precise by use of a calculated momentum
shift that depends on the electron's energy loss, $\omega$.
 A summary and some conclusions are presented in
Section~\ref{sec:summary}.

\section{Eikonal expansion for Klein-Gordon Waves}
\label{sec:eikonal_KG}

     For the Klein-Gordon equation with a
   Coulomb potential $V(r)$, one has
   \begin{equation}
  \Big( [ E - V(r)]^2 - {\bf p}^2 -m^2 \Big) \psi({\bf r}) = 0.
  \end{equation}
The potential may be a point-like Coulomb potential or a
potential that is derived from a finite charge distribution, such as that of a nucleus.
Writing the wave function in the form of a plane wave propagating
in the z-direction with wave number $k = \sqrt{E^2 - m^2}$ and a
complex phase shift $\bar{\chi}$,
\begin{equation}
\psi({\bf r}) = e^{i k z} e^{i \bar{\chi}({\bf r})},
\end{equation}
leads to the following equation for the phase shift,
\begin{equation}
[ E - V(r)]^2 - [k \hat{z} + \nabla \bar{\chi} -i\nabla ]\cdot[k
\hat{z} + \nabla \bar{\chi}] - m^2 = 0.
\label{eq:KG}
\end{equation}
Using the fact the $E^2 - k^2 - m^2 = 0$ and dividing by $2k$
leads to a differential equation for $\bar{\chi}$,
\begin{eqnarray}
 \frac{\partial \bar{\chi}}{\partial z}  =
-\frac{V}{v} + \frac{V^2}{2k} - \frac{(\nabla \bar{\chi} )^2}{2
k} + i \frac{\nabla^2 \bar{\chi}}{2k} , \label{eq:KG-eikonal-diff}
\end{eqnarray}
where $v = k/E \approx 1$ when $E>>m$.

 For outgoing-wave boundary
conditions and a potential that decreases faster than $1/r$, the
wave function must be a pure plane wave as $z \rightarrow
-\infty$ and a phase-shifted plane wave as $z \rightarrow
\infty$. For a Coulomb potential, the same boundary condition
is used with the understanding that the potential is cut off
at a large distance $r > \Lambda$.
For either case, the outgoing-wave eikonal phase vanishes as
$z \rightarrow -\infty$ and the
eikonal phase is found by integrating Eq.~(\ref{eq:KG-eikonal-diff})
as follows,
\begin{eqnarray}
\bar{\chi}^{(+)}({\bf r}) &=& -\frac{1}{v}\int_{- \infty}^{z} dz'
V(r')  \nonumber \\
&~&~~- \frac{1}{2k}\int_{- \infty}^{z} dz' \Biggr((\nabla'
\bar{\chi}^{(+)}({\bf r}'))^2 - V^2\Biggr) \nonumber \\
&~&~~+ \frac{i}{2k} \int_{- \infty}^{z} dz'{\nabla '}^2
\bar{\chi}^{(+)}({\bf r}').
 \label{eq:KG-eikonal}
\end{eqnarray}
A superscript $(+)$ denotes the outgoing-wave boundary condition.

The eikonal expansion is an iterative solution of
Eq.~(\ref{eq:KG-eikonal}) about the limit $k \rightarrow \infty$.
It initially takes the form
\begin{equation}
\bar{\chi}^{(+)} = \bar{\chi}_0^{(+)} + \bar{\chi}_1^{(+)} +
\bar{\chi}_2^{(+)} + \cdots,
\end{equation}
where the ``barred'' phase shifts are complex and their subscripts
denote the order of iteration. The lowest order term is
appropriate to the limit $k \rightarrow \infty$ and is obtained
from the first term on the right side of
Eq.~(\ref{eq:KG-eikonal}),
\begin{equation}
\bar{\chi}^{(+)}_0({\bf r}) = -\frac{1}{v}\int_{- \infty}^{z} dz'
V(r'),
 \label{eq:KG-chi_0}
\end{equation}
where $r' = \sqrt{z'^2 + b^2}$ and $b$ is the impact parameter.

 Higher order terms in the expansion are evaluated as follows.
 In first order, one
evaluates the left side of Eq.~(\ref{eq:KG-eikonal}) using
$\bar{\chi}^{(+)}_0 + \bar{\chi}^{(+)}_1$ and the right side
using $\bar{\chi}^{(+)}_0$.  After cancelling terms that are
equal because of Eq.~(\ref{eq:KG-chi_0}), the first-order term
$\bar{\chi}^{(+)}_1$ is obtained as
\begin{eqnarray}
\bar{\chi}^{(+)}_1({\bf r}) = - \frac{1}{2k}\int_{- \infty}^{z}
dz' \Biggr((\nabla' \bar{\chi}^{(+)}_0({\bf r}'))^2 - V^2\Biggr)
\nonumber \\
+ \frac{i}{2k} \int_{- \infty}^{z} dz'{\nabla '}^2
\bar{\chi}^{(+)}_0({\bf r}'). \label{eq:KG-chi_1}
\end{eqnarray}
In second order, one evaluates the left hand side of
Eq.~(\ref{eq:KG-eikonal}) using $\bar{\chi}^{(+)}_0 +
\bar{\chi}^{(+)}_1 + \bar{\chi}^{(+)}_2$ and the right hand side
using $\bar{\chi}^{(+)}_0 + \bar{\chi}^{(+)}_1$.  After cancelling
terms that are equal because of Eqs.~(\ref{eq:KG-chi_0}) and
(\ref{eq:KG-chi_1}), the second-order term is obtained as
\begin{eqnarray}
\bar{\chi}^{(+)}_2({\bf r}) = - \frac{1}{2k} \int_{- \infty}^{z}
dz'\Bigr[ 2\nabla ' \bar{\chi}^{(+)}_0({\bf r}')\cdot \nabla'
\bar{\chi}^{(+)}_1({\bf r}') \nonumber \\
+ (\nabla' \bar{\chi}^{(+)}_1({\bf r}'))^2 \Bigr] + \frac{i}{2k}
\int_{- \infty}^{z} dz' {\nabla '}^2
\bar{\chi}^{(+)}_1({\bf r}'). \label{eq:KG-chi_2} \nonumber \\
\end{eqnarray}
Higher-order terms are not  considered in this work.

    At the final stage of analysis, it is convenient to separate
the real and imaginary parts of the eikonal phase as follows,
\begin{eqnarray}
\bar{\chi}^{(+)}= \chi^{(+)} +  i \omega^{(+)}
\end{eqnarray}
and to expand each of these ``unbarred'' phases in systematic
powers of $1/k$,
\begin{eqnarray}
 \chi^{(+)} &=& \chi^{(+)}_0 + \chi^{(+)}_1 + \chi^{(+)}_2 + \cdots
\label{eq:KG-chi_exp(+)}
\nonumber \\
 \omega^{(+)} &=& \omega^{(+)} _1 + \omega^{(+)} _2 + \cdots .
\label{eq:KG-omega_exp(+)}
\end{eqnarray}
Here the subscripts denote the power of $1/k$ that is involved.
 Through second order, the
systematically ordered phases for the Klein-Gordon waves are,
\begin{eqnarray}
\chi^{(+)}_0({\bf r}) &=& -\frac{1}{v} \int_{- \infty}^{z} dz'
V(r')
\nonumber \\
\chi^{(+)}_1({\bf r}) &=& - \frac{1}{2k} \int_{- \infty}^{z} dz'
\Big((\nabla ' \chi^{(+)}_0({\bf r}'))^2 - V^2(r') \Big)
\nonumber \\
\chi^{(+)}_2({\bf r}) &=& -\frac{1}{2k} \int_{- \infty}^{z} dz'
\Bigr[ 2 \nabla ' \chi^{(+)}_0({\bf r}') \cdot \nabla '
\chi^{(+)}_1({\bf r}') \nonumber \\ &+&~~~~ \nabla^{'2}
\omega_1^{(+)}({\bf r}')\Bigr]
\nonumber \\
\omega^{(+)}_1({\bf r}) &=& \frac{1}{2k} \int_{- \infty}^{z} dz'
{\nabla '}^2 \chi^{(+)}_0({\bf r}')
\nonumber \\
\omega^{(+)}_2({\bf r}) &=& \frac{1}{2k} \int_{- \infty}^{z} dz'
\Big[ {\nabla '}^2 \chi^{(+)}_1({\bf r}')
\nonumber \\
&~& ~~-2\nabla'\chi_0^{(+)}({\bf r}') \cdot \nabla' \omega_1^{(+)}
({\bf r}') \Big] \label{eq:KG-phases}
\end{eqnarray}

For the Klein-Gordon wave function with outgoing-wave boundary
conditions, this gives
\begin{equation}
\psi^{(+)} ({\bf r}) = e^{i k z} e^{i\chi^{(+)}} e^{-
\omega^{(+)}},
\end{equation}
and one may work at various orders in the eikonal expansion by
truncating the expansions of Eq.~(\ref{eq:KG-phases}).

Incoming wave boundary conditions are appropriate for
final-state wave functions in matrix elements.  In that case the wave function
must be a pure plane wave as $z \rightarrow \infty$ and a
phase-shifted plane wave as $z \rightarrow -\infty$.  It is
written as
\begin{equation}
\psi^{(-)} ({\bf r}) = e^{i k z} e^{-i\bar{\chi}^{(-)}},
\end{equation}
which leads to
\begin{eqnarray}
 \frac{\partial \bar{\chi}^{(-)}}{\partial z}  =
\frac{V}{v} - \frac{V^2}{2k} + \frac{(\nabla \bar{\chi}^{(-)}
)^2}{2 k} + i \frac{\nabla^2 \bar{\chi}^{(-)}}{2k} .
\end{eqnarray}
Integration with incoming-wave boundary conditions produces
\begin{eqnarray}
\bar{\chi}^{(-)}({\bf r}) &=& -\frac{1}{v}\int_z^{ \infty} dz'
V(r')
\nonumber \\
&~& ~~- \frac{1}{2k}\int_z^{ \infty} dz' \Biggr((\nabla'
\bar{\chi}^{(-)}({\bf r}'))^2 - V^2\Biggr) \nonumber \\
&~&~~- \frac{i}{2k} \int_z^{\infty} dz'{\nabla '}^2
\bar{\chi}^{(-)}({\bf r}').
 \label{eq:KG-eikonal(-)}
\end{eqnarray}
As before, the eikonal expansion is an iterative solution of
Eq.~(\ref{eq:KG-eikonal(-)}) that produces an expansion about the
limit $k \rightarrow \infty$. It initially takes the form
\begin{equation}
\bar{\chi}^{(-)} = \bar{\chi}_0^{(-)} + \bar{\chi}_1^{(-)} +
\bar{\chi}_2^{(-)} + \cdots .
\end{equation}
 with the lowest order term being
\begin{equation}
\bar{\chi}^{(-)}_0({\bf r}) = -\frac{1}{v}\int_z^{ \infty} dz'
V(r').
 \label{eq:KG-chi_0(-)}
\end{equation}

 Higher order terms in the expansion are evaluated in the same manner
 as discussed above.  They are
\begin{eqnarray}
\bar{\chi}^{(-)}_1({\bf r}) &=& - \frac{1}{2k}\int_z^{ \infty} dz'
\Biggr((\nabla' \bar{\chi}^{(-)}_0({\bf r}'))^2 - V^2\Biggr)
\nonumber \\ &~&~~- \frac{i}{2k} \int_z^{ \infty} dz'{\nabla '}^2
\bar{\chi}^{(-)}_0({\bf r}'),
\nonumber \\
 \bar{\chi}^{(-)}_2({\bf r}) &=& -
\frac{1}{2k} \int_z^{\infty} dz'\Big[ 2\nabla '
\bar{\chi}^{(-)}_0({\bf r}')\cdot \nabla' \bar{\chi}^{(-)}_1({\bf
r}') \nonumber \\ &~&+ (\nabla' \bar{\chi}^{(-)}_1({\bf r}'))^2
\Big] - \frac{i}{2k} \int_z^{\infty} dz' {\nabla '}^2
\bar{\chi}^{(+)}_1({\bf r}').\nonumber \\
\end{eqnarray}

    Separating the real and imaginary parts of the eikonal phases as follows,
\begin{eqnarray}
\bar{\chi}^{(-)}= \chi^{(-)} -  i \omega^{(-)},
\label{eq:omega(-)}
\end{eqnarray}
and expanding each of these ``unbarred'' phases in systematic
powers of $1/k$, we have
\begin{eqnarray}
 \chi^{(-)} &=& \chi^{(-)}_0 + \chi^{(-)}_1 + \chi^{(-)}_2 + \cdots
\label{eq:KG-chi_exp(-)}
\nonumber \\
 \omega^{(-)} &=& \omega^{(-)} _1 + \omega^{(-)} _2 + \cdots .
\label{eq:KG-omega_exp(-)}
\end{eqnarray}
Such systematically ordered phases for the Klein-Gordon waves for
incoming wave boundary conditions may be obtained from
Eq.~(\ref{eq:KG-phases}) by replacing all integrations
$\int_{-\infty}^{z}$ by $\int _z^{\infty}$ and all superscripts
$(+)$ by $(-)$.  Because of the symmetry of the potential with
respect to inversion of $z$, the following relations hold for
each order $n$,
\begin{eqnarray}
\chi_n^{(-)}(z, {\bf b}) = \chi_n^{(+)}(-z, {\bf b}), \nonumber \\
\omega_n^{(-)}(z, {\bf b}) = \omega_n^{(+)}(-z, {\bf b}).
\label{eq:chi(-z)}
\end{eqnarray}
For the Klein-Gordon wave function with incoming wave boundary
conditions, this gives
\begin{equation}
\psi^{(-)} ({\bf r}) = e^{i k z} e^{-i\chi^{(-)}} e^{-
\omega^{(-)}},
\end{equation}
and one may work at various orders by truncating the expansions
for the eikonal phases.

 The imaginary part of the
eikonal phase produces the ``focusing factor'',
\begin{equation}
f^{KG(\pm)}({\bf r}) = e^{-\omega^{(\pm)}({\bf r})}.
\label{eq:KG-focus}
\end{equation}
In order to provide some insight into the focusing factor, it is
useful to consider a simple potential for which the eikonal
phases may be determined analytically.  That is done in
Appendix~\ref{app:analytical} and we find that at ${\bf r} = {\bf
0}$,
\begin{equation}
\omega_1^{(\pm)}({\bf 0}) = \frac{V(0)}{2kv} \approx
\frac{V(0)}{2E},
\end{equation}
where $V(0)$ is the potential at the origin.  Thus, the focusing
factor is approximately
\begin{equation}
f^{KG(\pm)}({\bf 0}) \approx 1 - V(0)/(2E).
\end{equation}

A matrix element for emission of a photon with momentum ${\bf q}$
and energy $\omega = E_i - E_f$ involves initial and final
momenta ${\bf k}_i$ and ${\bf k}_f$.  It takes the form
\begin{eqnarray}
&M&^{\mu}=\int d^3r \psi_{{\bf k}_f}^{(-)*}({\bf r}) j_{KG}^{\mu}
e^{-i {\bf q}\cdot {\bf r}} \psi_{{\bf k}_i}^{(+)}({\bf r}),
\nonumber \\
&~&=\int d^3r e^{i ({\bf Q} - {\bf q})\cdot {\bf r}} e^{i\chi}
 f^{KG(-)}_f({\bf r})f^{KG(+)}_i({\bf r}) j_{KG}^{\mu}
 \label{eq:KGME}
\end{eqnarray}
where ${\bf Q} = {\bf k}_i - {\bf k}_f$ and $\chi = \chi_f^{(-)} +
\chi_i^{(+)}$.  It should be noted that  $\chi^{(+)}_i$ and $\omega_i^{(+)}$
are obtained from Eqs.~(\ref{eq:KG-phases}) with the z-axis
parallel to initial momentum ${\bf k}_i$, while $\chi^{(-)}_f$
and $\omega_f^{(-)}$ are obtained from the same equations using
the outgoing-wave condition ($\int _z ^{\infty}$) with the
z-direction parallel to final momentum ${\bf k}_f$.

The conserved current for the Klein-Gordon equation also contains Coulomb
distortions, i.e.,
\begin{eqnarray}
&&j_{KG}^0 = \big[E_i + E_f - 2V(r)\big]/\sqrt{4E_iE_f},
\nonumber \\
&&{\bf j}_{KG} = \big[{\bf k}_i + {\bf k}_f +
\nabla \overline{\chi}_i^{(+)}({\bf r}) -
\nabla \overline{\chi}_f^{(-)}({\bf r})\big]/\sqrt{4E_iE_f},
\nonumber \\
\label{eq:KG_current}
 \end{eqnarray}
where phase-space factors for initial and final states have been included.
Current conservation holds because
\begin{eqnarray}
\Big(\omega j_{KG}^0 - {\bf q}\cdot {\bf j}_{KG}\Big)e^{-i{\bf q}\cdot {\bf r}} 
 \propto e^{-i{\bf q}\cdot {\bf r}} G^{-1}(E_i)-\nonumber \\
 G^{-1}(E_f)e^{-i{\bf q}\cdot {\bf r}} ,
\label{eq:WTIKG}
\end{eqnarray}
where $G^{-1}(E) = [E- V(r)]^2 - {\bf p}^2 -m^2$
annihilates the wave function at energy $E$.
Relative to the plane-wave current, the
Klein-Gordon current $j^0_{KG}$ contributes to the focusing effect a factor
equal to $(1 - V(r)/\bar{E})$ where $\bar{E} = {1
\over 2}(E_i+E_f)$ is the average energy.
 The initial-state wave
function provides a factor given by Eq.~(\ref{eq:KG-focus}),
i.e., $f_i^{D(+)} \approx (1 - V(0)/(2\bar{E}))$ and the final state wave
function provides a similar factor. The combination of these
factors at ${\bf r}=0$ is approximately equal to $(1 - V(0)/\bar{E})^2$.
That agrees with the result of Yennie et al. \cite{Yennie65} for
the Dirac wave function.

In the case of Dirac-Coulomb waves, the conserved electron
current is the Dirac matrix $\gamma^{\mu}$,
which it is not modified by the presence of a Coulomb potential. It
does not contain any focusing effects. The Dirac focusing
factors arise solely from the wave functions and are given in
Eq.~(\ref{eq:Dirac-focus}), which yields $f^{D(\pm)}\approx (1 -
V(0)/\bar{E})$ for each wave function. Thus, there is a
different focusing factor than for a Klein-Gordon wave function.
However, the combination of currents and wave functions produces
similar overall
focusing effects for the Klein-Gordon and Dirac
current matrix elements.  We note that the analysis of
Refs.~\cite{Aste04a,Aste04b} used the Dirac focusing factors with
the Klein-Gordon current, which produces one too many factors
of $1 - V(0)/\bar{E}$.

\section{\label{sec:Dirac} Eikonal expansion for Dirac-Coulomb waves }
\label{sec:eikonal_Dirac}

For the Dirac equation, the eikonal expansion is carried
out in two stages.
First we consider the Pauli spinor $u({\bf r})$ that
contains the two upper components of the Dirac wave function,
i.e.,
\begin{equation}
\psi({\bf r}) = \begin{pmatrix} u({\bf r}) \cr \ell({\bf
r})\end{pmatrix}.  \label{eq:psi}
\end{equation}
It follows from the Dirac equation that the Pauli spinor
$\ell({\bf r})$ that contains the two lower components may be
determined in a second stage from
\begin{equation}
\ell({\bf r}) = \frac{1}{E_2 - V}~\sigma \cdot {\bf p}~ u({\bf
r}), \label{eq:lowercomp}
\end{equation}
where $E_2 = E + m$.

\subsection{Upper component spinor }
\label{sec:upper-components}

Eliminating the lower component spinor from
the Dirac equation leads to the following equation for the
upper-component spinor
\begin{equation}
\Big( E_1 - V - \sigma \cdot {\bf p} \frac{1}{E_2 - V} \sigma
\cdot {\bf p} \Big)u({\bf r}), \label{eq:uofrDirac}
\end{equation}
where $E_1 = E - m$.  For electron scattering
it is generally the case that $E >> m$ and thus $E_1 \approx E_2
\approx E$.

For outgoing-wave boundary conditions, the Pauli spinor $u({\bf
r})$ is written in terms of a complex eikonal phase
$\bar{\chi}^{(+)}({\bf r})$ and a complex spin-dependent phase
$\bar{\gamma}^{(+)}({\bf r})$ as follows
\begin{equation}
u^{(+)}({\bf r}) = \left( 1 - \frac{V}{E_2}\right) ^{1/2} e^{ik z
} e^{i \bar{\chi}^{(+)}}e^{i \sigma_e \bar{\gamma}^{(+)}}.
\label{eq:chi+}
\end{equation}
The wave propagates in the $z$-direction and an impact vector
${\bf b}$ is defined as the part of ${\bf r}$ that is
perpendicular to the $\hat{z}$-direction, i.e., ${\bf b} =
\hat{z}\times ({\bf r} \times \hat{z})$.  Three orthogonal unit
vectors are : $\hat{z}$, $\hat{b} = {\bf b}/|{\bf b}|$ and
$\hat{e} = \hat{b} \times \hat{z}$. The spin matrix in the
eikonal phase is $\sigma_e = \sigma \cdot \hat{e}$. The factor
$(1 - V/E_2)^{1/2}$ is introduced in order to sum up terms that
otherwise arise in higher orders.

The eikonal expansion is developed in Appendix~\ref{app:dirac_eikonal_exp}.
The result is that the eikonal phases are expanded in a systematic fashion
in powers of $1/k$ as follows.
\begin{eqnarray}
 \chi^{(+)} &=& \chi^{(+)}_0 + \chi^{(+)}_1 + \chi^{(+)}_2 + \cdots
\label{eq:chi_exp}
\nonumber \\
 \omega^{(+)} &=& ~~~~~~~~~~\omega^{(+)} _1 + \omega^{(+)} _2 + \cdots
\label{eq:omega_exp}
\nonumber \\
 \gamma^{(+)} &=& ~~~~~~~~~~\gamma^{(+)}_1 + \gamma^{(+)}_2 + \cdots
\label{eq:gamma_exp}
\nonumber \\
\delta^{(+)} &=& ~~~~~~~~~~~~~~~~~~~~\delta^{(+)}_2 +  \cdots ,
\end{eqnarray}
where the subscript of each term denotes the power of $1/k$ that
is involved. The systematically ordered phases for the Dirac
equation are as follows.
\begin{eqnarray}
\chi^{(+)}_0({\bf r}) &=& -\frac{1}{v} \int_{- \infty}^{z} dz'
V(r'),
\nonumber \\
\chi^{(+)}_1({\bf r}) &=& - \frac{1}{2k} \int_{- \infty}^{z} dz'
\Big((\nabla ' \chi^{(+)}_0({\bf r}'))^2 - V^2(r') \Big),
\nonumber \\
\chi^{(+)}_2({\bf r}) &=& -\frac{1}{2k} \int_{- \infty}^{z} dz'
\Bigr[( 2 \nabla ' \chi^{(+)}_0({\bf r}') \cdot \nabla '
\chi^{(+)}_1({\bf r}') \nonumber \\ &+&  \nabla^{'2}
\omega_1^{(+)}({\bf r}')\Big]  - \frac{1}{4kE_2} {\nabla}^2
\chi_0^{(+)},
\nonumber \\
\omega^{(+)}_1({\bf r}) &=& \frac{1}{2k} \int_{- \infty}^{z} dz'
{\nabla '}^2 \chi^{(+)}_0({\bf r}'),
\nonumber \\
\omega^{(+)}_2({\bf r}) &=& \frac{1}{2k} \int_{- \infty}^{z} dz'
\Bigr[ {\nabla '}^2 \chi^{(+)}_1({\bf r}') \nonumber \\ &-&
2\nabla'\chi_0^{(+)}({\bf r}') \cdot \nabla' \omega_1^{(+)} ({\bf
r}') \Bigr],
\nonumber \\
\gamma^{(+)}_1({\bf r}) &=& -\frac{1}{2k}  \int_{- \infty}^{z} dz'
\frac{\partial V(r)}{\partial b}
\nonumber \\
\gamma^{(+)}_2({\bf r}) &=& -\frac{1}{2k} \int_{- \infty}^{z} dz'
\Biggr( 2\nabla'\chi_0^{(+)}({\bf r}') \cdot \nabla'
\gamma_1^{(+)} ({\bf r}') \nonumber \\ &-& \frac{1}{E}
\frac{\partial V}{\partial z} \frac{\partial \chi_0^{(+)}}
{\partial b} \Biggr)
\nonumber \\
\delta^{(+)}_2({\bf r}) &=& \frac{1}{2k}  \int_{- \infty}^{z} dz'
\Biggr( \nabla^{' 2} - \frac{1}{b^2} \Biggr) \gamma_1^{(+)} ({\bf r}').
\label{eq:eikonal_exp}
\end{eqnarray}
Results for $\chi_0^{(+)}$, $\chi_1^{(+)}$, $\omega_1^{(+)}$ and
$\omega_2^{(+)}$ are the same as for the Klein-Gordon case. The
last term in the result for $\chi_2^{(+)}$ is not present in the
Klein-Gordon case.

Spin-dependent phases can be shown to be simply related to the
spin-independent ones as follows,
\begin{eqnarray}
\gamma_1^{(+)} + \gamma_1^{(+)} + i \delta_2^{(+)} =
\frac{1}{E_2-V} \frac{\partial}{\partial b} \Big[ \chi_0^{(+)} +
\chi_1^{(+)} + i \omega_1^{(+)} \Big].\nonumber
\end{eqnarray}
This connection holds to order $1/k^2$.

 The upper-component spinor of the Dirac wave function
for helicity $\lambda$ and outgoing-wave boundary conditions is
given by
\begin{equation}
u^{(+)}_{\lambda} = \Bigr( 1 - {V \over E_2}\Bigr)^{1/2} e^{i k
z} e^{i\chi^{(+)}} e^{-\omega^{(+)}} e^{i \sigma_e
\bar{\gamma}_i^{(+)}} \xi_{\lambda}, \label{eq:u+ofr}
\end{equation}
where $\xi_{\lambda}$ is a helicity eigenstate.  One may work at
various orders of the eikonal expansion by truncating the
expansions of Eq.~(\ref{eq:chi_exp}).

The upper-component spinor for helicity $\lambda$ and
incoming-wave boundary conditions is given by
\begin{equation}
u^{(-)}_{\lambda} =  \Bigr(1 - {V\over E_2}\Bigr)^{1/2} e^{i k z}
e^{-i\chi^{(-)}} e^{-\omega^{(-)}} e^{-i \sigma_e
\bar{\gamma}^{(-)}}\xi_{\lambda}, \label{eq:u-ofr}
\end{equation}
where
\begin{eqnarray}
\overline{\gamma}^{(-)} &=& \gamma^{(-)} - i \delta^{(-)} \nonumber \\
&=&  \gamma_1^{(-)} + \gamma_2^{(-)} \cdots - i ( \delta_2^{(-)} +
\cdots),
\end{eqnarray}
 with phases obtained from Eqs.~(\ref{eq:chi_exp}) and (\ref{eq:eikonal_exp}) by
replacing the superscripts $(+)$ by $(-)$ and the integration
ranges $\int_{-\infty}^z$ by $\int_z^{\infty}$. Alternatively, the
symmetry of Eq.~(\ref{eq:chi(-z)}) may be used.

\subsection{Lower-component spinors}
\label{sec:lower-components}

Equation~(\ref{eq:lowercomp}) specifies the lower-component
spinor in terms of the upper-component spinor.  The connection is
very simple in the limit $m \rightarrow 0$ when helicity
eigenstates are used. We find
\begin{eqnarray}
 \ell_{\lambda} ({\bf r})  =  2 \lambda u_{\lambda} ({\bf r})
\label{eq:lofr}
\end{eqnarray}
where $\lambda = \pm {1 \over 2}$.  This holds generally in the
$m=0$ limit as the following analysis shows.

For $m=0$, we have $E_1 = E_2 = E$. If we define
$\tilde{u}_{\lambda} = \Big( 1 - V/E\Big)^{1/2} u_{\lambda}({\bf
r})$, then Eq.~(\ref{eq:uofrDirac}) may be written as
\begin{eqnarray}
 \tilde{u}_{\lambda} ({\bf r}) = h^2 \tilde{u}_{\lambda},
\end{eqnarray}
where for $E > |V|$, $h$ is the hermitian operator
\begin{eqnarray}
 h \equiv \frac{1}{\sqrt{E - V}}~\sigma \cdot {\bf p}~
\frac{1}{\sqrt{E - V} }.
 \end{eqnarray}
Thus, $\tilde{u}_{\lambda}$ is an eigenfunction of the hermitian
operator $h^2$ with eigenvalue +1. It must then also be an
eigenfunction of $h$ with eigenvalues $\pm 1$, that is,
\begin{equation}
 h \tilde{u}_{\lambda} = \pm  \tilde{u}_{\lambda}.
\end{equation}
However, because of the Dirac equation it follows that
\begin{equation}
 h \tilde{u}_{\lambda} = \tilde{\ell}_{\lambda},
\end{equation}
where  $\tilde{\ell}_{\lambda} = \Big( 1 - V/E\Big)^{1/2} \ell_{\lambda}$.
These two equations require that
\begin{equation}
  \tilde{\ell}_{\lambda} = 2 \lambda \tilde{u}_{\lambda}
\end{equation}
where $\lambda = \pm{1 \over 2}$, which proves
Eq.~(\ref{eq:lofr}). Clearly, $h$ is the helicity operator for
the $m=0$ Dirac equation. Conservation of helicity is a
well-known result for the limit $m\rightarrow 0$ when the
interaction is a vector current, or any single component of a
vector interaction as for the Coulomb interaction.  In the limit
as $V \rightarrow 0$, $h$ becomes the usual helicity operator for
a plane-wave state.

For a nonzero electron mass, a similar but approximate analysis
may be performed to show that
\begin{eqnarray}
\ell_{\lambda}({\bf r}) &\approx& 2 \lambda \Biggr( \frac{E_1 -
\l<V\r>}{E_2 -
\l<V\r>}\Biggr)^{1/2} u_{\lambda} ({\bf r}), \nonumber \\
& \approx & 2 \lambda \Bigr( 1 - \frac{m}{E_2 - \l<V\r>} \Bigr)
u_{\lambda}({\bf r}),
\end{eqnarray}
where $\l<V\r>$ is the average potential.  The correction term is
about one part per thousand for a 500 MeV electron.  It will be
neglected in this work.


\subsection{Convergence of eikonal expansion}
\label{sec:convergence}

Convergence of the eikonal expansion is discussed for scattering
of a 500 MeV electron by $^{208}Pb$. Given that the electron mass
is $m = .511 MeV$, it follows that $k \approx E$ and $v \approx
1$, both within a part per million.  The Coulomb potential is
approximately $V(0) = 25$ MeV at the center of the nucleus.  The
leading order eikonal phase is of order $2R V(0)$, where $R$ is
the mean radius and the factor $2R$ provides an estimate of the
integral over $z$.  If the charge distribution is approximated as
constant within a sphere of radius $R$, then $V(0) = {3 \over 2}
Z\alpha/R$, so we expect that $\chi^{(+)}_0 \approx 3 Z \alpha$.
For $^{208}Pb$, this gives $\chi^{(+)}_0 \approx 2$.  The eikonal
expansion introduces corrections that involve the nondimensional
ratio $V/E \approx .05$, so we expect $\chi^{(+)}_2 \approx .0025
\chi^{(+)}_0 \approx .005$. The eikonal expansion is not
convergent but is asymptotic, meaning that the error should be
bounded by the first neglected term. When terms up to second
order are kept, the error is of order $\chi^{(+)}_3 \approx
.000125 \chi^{(+)}_0 \approx .00025$ in the example discussed
here.  There is a second expansion parameter involved in terms
like $\omega_1$ or $\omega_2$, namely $1/ka$, where $a$ is a
length parameter that characterizes derivatives of the potential,
i.e., $|\nabla V(r)| \approx 1/a |V(r)|$. Provided that the
potential is sufficiently smooth, these corrections also are
small. Thus, the expansion can produce accurate wave functions
for electron scattering when the energy is sufficiently high and
the potential is sufficiently smooth. One must be aware that the
eikonal wave functions omit the wave bending effects that arise
from exact solutions and are less accurate far from the nucleus.
However, the eikonal wave functions can provide an accurate
description of Coulomb distortion effects in the region where the
nuclear charge density is nonzero.

It is possible to reduce the terms in the eikonal expansion to
analytical forms for the potential
\begin{equation}
V(r) = - \frac{\alpha Z}{\sqrt{r^2 + R^2}}, \label{eq:Vofr}
\end{equation}
where $\alpha = e^2/(\hbar c)$, Z is the nuclear charge and $R$ is
a length parameter.  This Coulombic potential corresponds to a
charge density
\begin{equation}
\rho(r) = \frac{3 Z e}{4 \pi} \frac{R^2}{(r^2 + R^2)^{5/2}}.
\end{equation}
The analytical results are given in Appendix~\ref{app:analytical}.

Figure~\ref{fig:EikPhases} shows the eikonal phases for a charge
$Z=100$, electron energy $E=200$ MeV and radius parameter $R=2$ fermi.
These parameters are chosen in order to make the corrections
visible.  The corrections are much smaller for a 500 MeV electron
and smaller nuclear charge.
\begin{figure}[h]
\includegraphics[width=7cm,bb= 20 40 480 530,clip]{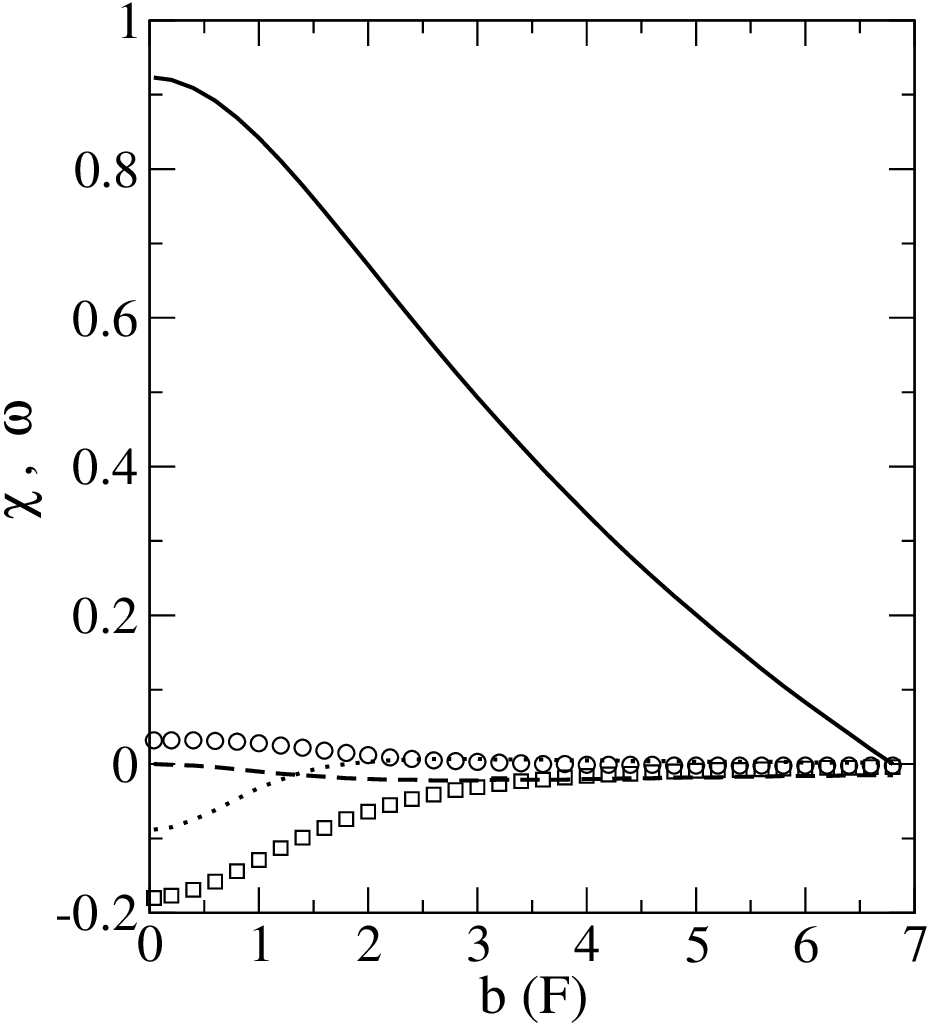}
\caption{Eikonal phases at z=0 versus impact parameter: solid line
shows $\chi_0$, dashed line shows $\chi_1$, dotted line shows
$\chi_2$, rectangles show $\omega_1$ and ovals show $\omega_2$. A
constant has been added to $\chi_0$ such that it vanishes at $b =
3.5R$.  Phases are shown for Z=100, E=200 MeV and R=2 fermi.
}\label{fig:EikPhases}
\end{figure}

\subsection{Focusing factors}
\label{sec:focusing-factors}

As noted above, the focusing factors are important when Coulomb
distorted waves are used.  Ignoring for the moment the lower
component spinors, we consider a current matrix element for
emission of a photon of energy $\omega = E_i - E_f$ and momentum
${\bf q}$ using upper-component spinors corresponding to initial
momentum ${\bf k}_i$, initial helicity $\lambda_i$, final
momentum ${\bf k}_f$ and final helicity $\lambda_f$. The momentum
transfer is ${\bf Q} = {\bf k}_i - {\bf k}_f$ and the current
matrix element is
\begin{eqnarray}
&&M^{\mu} =  \int d^3ru_{\lambda_f}^{(-)*}({\bf
r})\gamma^{\mu}e^{-i
{\bf q}\cdot{\bf r}} u_{\lambda_i}^{(+)}({\bf r}) \nonumber \\
&&= \int d^3r \xi^{\dag}_{\lambda_f}
  e^{i({\bf Q} - {\bf q})\cdot {\bf r}} e^{ i \chi }
  e^{i\sigma_{e_f} \bar{\gamma}_f^{(-)*}} f^{D(-)}_f\gamma^{\mu}
f^{D(+)}_i \times \nonumber \\
&&e^{i \sigma_{e_i}
\bar{\gamma}_i^{(+)}} \xi_{\lambda_i },
\label{eq:ufbarOui}
\end{eqnarray}
where $\chi = \chi_f^{(-)}({\bf r}) + \chi_i^{(+)}({\bf r})$ and
the Dirac focusing factors are defined by
\begin{eqnarray}
  f^{D(+)}_i({\bf r}) &=& \Biggr( 1 - \frac{V}{E_{2i}}\Biggr)^{1/2}
  e^{-\omega_i^{(+)}}, \nonumber \\ f^{D(-)}_f({\bf r}) &=&
  \Biggr( 1 - \frac{V}{E_{2f}}\Biggr)^{1/2} e^{-\omega_f^{(-)}}.
  \label{eq:Dirac-focus}
  \end{eqnarray}
Note that  $\chi^{(+)}_i$, $\omega_i^{(+)}$, and
$\bar{\gamma}_i^{(+)} = \gamma_i^{(+)} + i \delta_i^{(+)}$ are
obtained from Eqs.~(\ref{eq:chi_exp}) to (\ref{eq:eikonal_exp})
with the z-axis parallel to initial momentum ${\bf k}_i$, while
$\chi^{(-)}_f$, $\omega_i^{(-)}$, and $\bar{\gamma}_i^{(-)} =
\gamma_i^{(-)} - i \delta_i^{(-)}$ ,  are obtained from the same
equations using the outgoing-wave condition ($\int _z ^{\infty}$)
with the z-direction parallel to final momentum ${\bf k}_f$.

Focusing factors differ from those appropriate to a Klein-Gordon
wave function by the factor $( 1 - V/E_{2i})^{1/2}$ as may be seen
by comparing Eqs.~(\ref{eq:KG-focus}) and (\ref{eq:Dirac-focus}).
The $e^{- \omega_i^{(+)}} \approx 1 - V(0)/(2E_i)$ factor is the
same to leading order for a Klein-Gordon and a Dirac wave
function. Combining this part of the eikonal correction with the
$(1 - V/E_{2i})^{1/2}$ yields a focusing factor $f^{D(+)}_i \approx 1 -
V/E_i$ in the Dirac wave function, thus reproducing at $r=0$ the
expected factor $1 - V(0)/E_i$ that has been derived by Yennie,
Boos and Ravenhall~\cite{Yennie65} based on a WKB analysis of the
Dirac-Coulomb wave function.
 A similar result
holds for the final-state focusing factor, $f^{D(-)}_f$, which is
approximately $1 -V/E_f$.  Thus, the overall focusing effect in the
matrix element is
approximately equal to $(1 -V(0)/E_f)(1 -V(0)/E_i)$, which is
similar to the overall effect in the Klein-Gordan matrix element,
Eq.~(\ref{eq:KGME}).
 The reason that
focusing factors did not appear in early analyses based on the
eikonal approximation is because they arise from corrections to
the eikonal approximation.

    In passing, we note that the Glauber approximation is
obtained when the eikonal phases for initial and final states are
evaluated using for each a z-axis parallel to the average
momentum, ${1\over 2}({\bf k}_i + {\bf k}_f)$, and only the
leading-order phases, $\chi^{(+)}_0$ and $\chi^{(-)}_0$, are
retained. This approximation omits the focusing factors.

\section{Electron current matrix elements}
\label{sec:currentME}
Electron scattering involves the current matrix element
\begin{equation}
 j_e^{\mu} = \int d^3 r \bar{\Psi}_{{\bf k}_f \lambda_f}^{(-)} ({\bf r})
\gamma^{\mu}e^{-i{\bf q}\cdot {\bf r}} \Psi_{{\bf k}_i
\lambda_i}^{(+)} ({\bf r}),
\end{equation}
where ${\bf q}$ is the three-momentum of a photon emitted at point
${\bf r}$, $\omega = E_i - E_f$ is the energy of the photon and
\begin{equation}
\Psi_{{\bf k}_i \lambda_i}^{(+)} ({\bf r}) =  \frac{1}{\sqrt{2}}
\begin{pmatrix} u_{\lambda_i}^{(+)}({\bf r}) \cr
\ell_{\lambda_i}^{(+)}({\bf r}) \end{pmatrix} ,
\end{equation}
is a Dirac-Coulomb wave with a normalization factor $1/\sqrt{2}$
included in order that it reduces as $V \rightarrow 0$ to the
$m=0$ plane-wave spinor,
\begin{equation}
u_{\lambda_i}({\bf k}_i) = \frac{1}{\sqrt{2}}
\begin{pmatrix} 1 \cr \sigma \cdot \hat{k}_i  \end{pmatrix} \xi_{\lambda_i}
e^{i {\bf k}_i\cdot {\bf r}}.
\end{equation}
Using lower-component spinors from Eq.~(\ref{eq:lofr}) and the
conventions of Bjorken and Drell~\cite{Bjorken65} for the
$\gamma^{\mu}$ matrices, one readily finds that the matrix
element $j_e^0$ involves the overall factor $1 + 4 \lambda_f
\lambda_i = 2 \delta_{\lambda_f \lambda_i}$ and the matrix
element ${\bf j}_{e}$ involves the overall factor $2\lambda_f +
2\lambda_i = (2 \lambda_i)2 \delta_{\lambda_f \lambda_i}$. Thus,
helicity is conserved as it must be in the $m = 0$ limit. The
electron current matrix elements are further reduced by use of the
convention of Kubis~\cite{Kubis72} for the helicity matrix
elements between initial and final states, leading to
\begin{equation}
 j_{e}^{\mu}   = \delta_{\lambda_f \lambda_i} \int d^3 r
e^{i ({\bf Q}-{\bf q})\cdot {\bf r}} e^{i \chi} f^{D(-)}_f({\bf r})
f^{D(+)}_i({\bf r}) h_e^{\mu}({\bf r})
\end{equation}
where $h_e^{\mu}({\bf r})$ is a four-vector of helicity matrix
elements,
\begin{eqnarray}
 \big\{h_e^{0}, {\bf h}_e \big\}  &=& \xi^{\dag}_{\lambda_f}(\theta_e) e^{i
\sigma_{e_f}\bar{\gamma}_f^{(-)*}} \big\{ 1, \vec{\sigma} \big\}
e^{i \sigma_{e_i}\bar{\gamma}_i^{(-)}} \xi_{\lambda_i}
\nonumber \\
 &=&  \xi^{\dag}_{\lambda_f}(\theta_e) \Big[cos\bar{\gamma}_f^{(-)*} +
i \sigma_{e_f}sin\bar{\gamma}_f^{(-)*} \Big]  \{ 1, \vec{\sigma}
\}
\nonumber \\
&& \Big[cos\bar{\gamma}_i^{(+)} + i
\sigma_{e_i}sin\bar{\gamma}_i^{(+)} \Big] \xi_{\lambda_i}.
\label{eq:J0}
\end{eqnarray}
Here $\xi^{\dag}_{\lambda_f}(\theta_e) = \xi^{\dag}_{\lambda_f}
e^{i \sigma_y \theta_e/2}$ is the helicity eigenstate for the
outgoing electron and $\theta_e$ is the scattering angle of the
electron. We find that
\begin{eqnarray}
i \sigma_{e_i} \xi_{\lambda_i} &=& 2\lambda_i e^{2 i \lambda_i
\phi_i} \xi_{-\lambda_i} , \nonumber \\
\xi^{\dag}_{\lambda_f}(\theta_e) i\sigma_{e_f} &=& -2 \lambda_f
e^{-2 i \lambda_f \phi_f}\xi^{\dag}_{-\lambda_f}(\theta_e).
\end{eqnarray}
Angle $\phi_i$ is the polar angle of initial state impact
parameter, i.e.,  ${\bf b}_i = cos\phi_i \hat{x} +
sin\phi_i\hat{y}$ and $\phi_f$ is the polar angle of the final
state impact parameter, where initial and final momenta are in
the $xz$-plane: ${\bf k}_i = k_i \hat{z}$ and
 ${\bf k}_f = k_f[
cos\theta_e \hat{z} + sin\theta_e\hat{x}]$. Moreover the required
helicity matrix elements are
\begin{eqnarray}
 \xi^{\dag}_{\lambda_f}(\theta_e)\xi_{\lambda_i} =
\delta_{\lambda_f \lambda_i} cos{1 \over 2}\theta_e +(\lambda_f -
\lambda_i) sin{1 \over 2}\theta_e
\nonumber \\
\xi^{\dag}_{\lambda_f}(\theta_e)~ \vec{\sigma}~ \xi_{\lambda_i}=
(\lambda_f + \lambda_i) \Big( \hat{e}_{2 \lambda_i} sin{1 \over
2}\theta_e + \hat{e}_z cos{1 \over 2}\theta_e \Big) \nonumber \\
+ |\lambda_f - \lambda_i| \Big( \hat{e}_{2 \lambda_i} cos{1 \over
2}\theta_e - \hat{e}_z  sin{1 \over 2}\theta_e \Big)
\end{eqnarray}
where $\hat{e}_{2 \lambda} = \hat{e}_x + (2 \lambda) i \hat{e}_y$.

Combining these parts produces the required helicity matrix
elements for the components of the current (plane-wave values are
shown following the arrows),
\begin{eqnarray}
h_e^0  &=&  A_{2 \lambda_i} cos{1 \over 2}\theta_e   + C_{2
\lambda_i} sin{1\over 2}\theta_e
\longrightarrow cos{1\over 2}\theta_e ,\nonumber \\
h_e^x  &=&   B_{2 \lambda_i} sin {1 \over 2}\theta_e  + D_{2
\lambda_i} cos{1 \over 2}\theta_e
\longrightarrow sin{1\over 2}\theta_e ,\nonumber \\
h_e^y&=&   2 i \lambda_i  \Bigr(A_{2 \lambda_i} sin{1 \over
2}\theta_e
  - C_{2 \lambda_i} cos{1 \over 2}\theta_e \Bigr)
\longrightarrow 2 i \lambda_i sin{1\over 2}\theta_e ,\nonumber \\
h_e^{z} &=&   B_{2 \lambda_i} cos{1 \over 2}\theta_e  -
D_{2 \lambda_i} sin {1 \over 2}\theta_e \longrightarrow cos{1\over 2}\theta_e ,\nonumber \\
 \label{eq:helicityfacs}
\end{eqnarray}
where the spin-orbit parts of the eikonal phases enter the
helicity matrix elements in the following four combinations
\begin{eqnarray}
A_{2 \lambda_i}\equiv cos \bar{\gamma}_f^{(-)*} cos
\bar{\gamma}_i^{(+)} - sin\bar{\gamma}_i^{(+)}
sin\bar{\gamma}_f^{(-)*} e^{2i \lambda_i (\phi_i - \phi_f) }, \nonumber \\
B_{2 \lambda_i} \equiv  cos \bar{\gamma}_f^{(-)*} cos
\bar{\gamma}_i^{(+)}+
sin\bar{\gamma}_i^{(+)}sin\bar{\gamma}_f^{(-)*} e^{2i \lambda_i (\phi_i - \phi_f) }, \nonumber \\
C_{2 \lambda_i} \equiv cos
\bar{\gamma}_f^{(-)*}sin\bar{\gamma}_i^{(+)}e^{2i \lambda_i
\phi_i} + cos\bar{\gamma}_i^{(+)} sin\bar{\gamma}_f^{(-)*}
e^{-2i \lambda_i \phi_f },  \nonumber \\
D_{2 \lambda_i} \equiv  cos \bar{\gamma}_f^{(-)*}
sin\bar{\gamma}_i^{(+)}e^{2i \lambda_i \phi_i} - cos
\bar{\gamma}_i^{(+)} sin\bar{\gamma}_f^{(-)*} e^{-2i \lambda_i
\phi_f }. \nonumber \\  \label{eq:ABCD}
\end{eqnarray}
Spin-orbit phases are of order $1/k$ so the corrections that they
produce are less important at higher electron energy.
The current matrix element given above is a general form
expressed in terms of the complex phase $\overline{\gamma}^{(\pm)} =
\gamma^{(\pm)} \mp i \delta^{(\pm)}$ and based on the zero-mass
limit for the lower components of the Dirac wave function. One
may work at various orders of approximation by using the results
of Section~\ref{sec:eikonal_Dirac} for the eikonal phases. The leading
order contributions are from $\gamma_1^{(\pm)}$, which is real.

   Some insight into the helicity matrix elements can be obtained
by evaluating them for forward scattering and backward
scattering.  For forward scattering, the impact parameters and
azimuthal angles for initial- and final-state eikonal phases are
equal, i.e., $b_f = b_i$ and $\phi_f = \phi_i$. Evaluating the
expression of Eq.~(\ref{eq:ABCD}) one finds the following simpler
forms,
\begin{eqnarray}
A_{2 \lambda_i} = cos(\bar{\gamma}_f^{(-)*}+
\bar{\gamma}_i^{(+)}), \nonumber \\
B_{2 \lambda_i} = cos(\bar{\gamma}_i^{(+)}- \bar{\gamma}_f^{(-)*}), \nonumber \\
 C_{2 \lambda_i}= \pm sin(
\bar{\gamma}_f^{(-)*}+\bar{\gamma}_i^{(+)}),  \nonumber \\
D_{2\lambda_i} =  \pm sin(\bar{\gamma}_f^{(-)*} -
\bar{\gamma}_i^{(+)}), \label{eq:phi=0andpi}
\end{eqnarray}
where the upper sign applies for $\phi_i=\phi_f =0$ and the lower
one for $\phi_i =\phi_f = \pi$. The helicity matrix elements for
the same values of $\phi_i$ are
\begin{eqnarray}
h_e^0 \rightarrow cos\Big({1\over 2}\theta_e \mp(
\bar{\gamma}_i^{(+)} + \bar{\gamma}_f^{(-)*}) \Big),
\nonumber \\
h_e^x \rightarrow sin\Big({1\over 2}\theta_e \pm
(\bar{\gamma}_i^{(+)} - \bar{\gamma}_f^{(-)*} ) \Big),
\nonumber \\
h_e^y \rightarrow (2 i \lambda_i) sin\Big({1\over 2}\theta_e
\mp(\bar{\gamma}_i^{(+)} + \bar{\gamma}_f^{(-)*}) \Big),
\nonumber \\
h_e^z \rightarrow cos \Big({1\over 2}\theta_e \pm
(\bar{\gamma}_i^{(+)} - \bar{\gamma}_f^{(-)*} ) \Big).
\label{eq:helicity-ME}
\end{eqnarray}
Note that the helicity matrix elements are similar to the
plane-wave matrix elements except for a shift of the electron
scattering angle.  The shift depends upon the azimuthal angle
$\phi_i$ and as indicated by the $\pm$ signs, the shifts at
$\phi_i = \pi$ are opposite to those at $\phi_i = 0$.
Cancellations are expected in the integration over $\phi_i$. These
shifts that occur due to the spin-orbit interaction affect the
longitudinal and transverse parts of the current in somewhat
different ways.  They may provide interesting insight into the
accuracy with which one may make the L/T separation in the
presence of Coulomb corrections.  However, the required
numerical evaluation
is beyond the scope of this paper.

\section{Quasi-elastic electron scattering by nuclei}
\label{sec:quasi-elastic}

     In this section we consider quasi-elastic scattering of electrons
     by nuclei but only taking into account the longitudinal
     current and the Coulomb corrections that arise
     from spin-independent terms in the eikonal expansion.
     The relevant matrix element involves one-photon exchange
     between the electron and a nucleon in the nucleus and the
     cross section takes a well-known form,
\begin{eqnarray}
\frac{d \sigma}{d\Omega_f dE_f} = \int d\Omega_p \frac{4 \alpha^2
}{(2\pi)^5 }\overline{\left|{\cal M} \right|^2} E_f^2 p E_p,
\label{eq:dsigmadomega}
\end{eqnarray}
where $p$ is the momentum of the knocked-out nucleon and $E_p =
\sqrt{M^2 + p^2}$ is its energy.  The bar denotes an average over
initial helicities and a sum over final helicities.  The
quasi-elastic matrix element is
\begin{eqnarray}
{\cal M} = \delta_{\lambda_f\lambda_i} \int d^3r \int
\frac{d^3q}{(2\pi)^2} e^{i ({\bf
Q}-{\bf q})\cdot {\bf r}} e^{i\chi({\bf r})}\times \nonumber \\
f^{D(-)}_f({\bf r})f^{D(+)}_i({\bf r}) h_e ^{\mu}({\bf r})
 \Biggr( \frac{1}{{\bf q}^2 -
\omega^2}\Biggr)  J^N_{\mu}({\bf q},{\bf p}).
 \label{eq:Mtotal}
\end{eqnarray}
In the plane-wave impulse approximation (PWIA), Coulomb
distortion of the electron waves is neglected so the integration
over ${\bf r}$ produces $\delta^{(3)}({\bf q} - {\bf Q})$. The
matrix element simplifies to
\begin{equation}
{\cal M}^{PWIA} = \delta_{\lambda_f \lambda_i}
\frac{h_{PWIA}^{\mu} J_{N \mu}({\bf Q},{\bf p})}{Q^2},
\label{eq:M_PWIA}
\end{equation}
where $h_{PWIA}^{\mu}$ denotes the helicity factors shown in
Eq.~(\ref{eq:helicityfacs}) after the arrows. The PWIA cross
section may be expressed in terms of longitudinal and transverse
response functions, $R_L$ and $R_T$, as follows
\begin{equation}
\frac{d \sigma}{d\Omega_f dE_f} = \sigma_{{\rm Mott}} \Biggr\{
\frac{Q^4}{{\bf Q}^4} R_L + \frac{Q^2}{2{\bf
Q}^2}\frac{1}{\epsilon}~ R_T\Biggr\},
\end{equation}
where
\begin{equation}
\sigma_{{\rm Mott}} = \frac{4\alpha^2 E_f^2 cos^2{\theta_e\over
2}}{Q^4},
\end{equation}
and
\begin{equation}
\epsilon = \Biggr[ 1 + \frac{2 {\bf Q}^2}{Q^2} tan^2 {\theta_e
\over 2} \Biggr]^{-1}.
\end{equation}

With Coulomb corrections included, the longitudinal matrix element
of interest must take a gauge invariant form.
This requires that the electron current must be conserved in the sense
that
\begin{eqnarray}
\int d^3r \Psi_{k_f}^{(-)*}({\bf r})
\Big(\omega j_e^0 - {\bf q}\cdot {\bf j}_e\Big)
e^{-i{\bf q}\cdot {\bf r}} \Psi_{k_i}^{(+)}({\bf r}) = 0,
\end{eqnarray}
and that the nuclear current should separately be conserved,
\begin{eqnarray}
\int d^3r \Psi_{p}^{(-)*}({\bf r})e^{i{\bf q}\cdot {\bf r}}
\Big( \omega J_N^0 - {\bf q}\cdot {\bf J}_N\Big)
\psi({\bf r}) = 0,
\label{eq:currcons}
\end{eqnarray}
where ${\bf q}$ is the photon three momentum.  With
Coulomb distorted waves, the photon momentum ${\bf q}$ differs
from the electron's momentum transfer ${\bf Q} = {\bf k}_i - {\bf
k}_f$ and the longitudinal current is defined with respect to the
direction of the photon that is exchanged, not with respect to
the difference of asymptotic electron momenta.
Current conservation follows because the current
obeys a Ward identity similar in form to Eq.~(\ref{eq:WTIKG}).

Owing to current conservation, the
longitudinal current matrix element can be simplified as follows,
\begin{eqnarray}
j_e^0 J_N^0 - (\hat{q}\cdot {\bf j}^e) (\hat{q}\cdot {\bf J}_N) =
j_e^0 J_N^0\Biggr( 1 - \frac{\omega^2}{{\bf q}^2}\biggr),
\label{eq:long_ME}
\end{eqnarray}
which holds either for the Klein-Gordon or Dirac case.

When the Dirac equation is used for the electron's
 Coulomb distorted waves,
the matrix element due to the longitudinal current is
\begin{eqnarray}
{\cal M}_L
 =&=& \delta_{\lambda_f\lambda_i}\int d^3r \int \frac{d^3q}{(2\pi)^2} e^{i ({\bf
Q}-{\bf q})\cdot {\bf r}} e^{i\chi({\bf r})}\times \nonumber \\
& &f^{D(-)}_f({\bf r})f^{D(+)}_i({\bf r}) h_e ^{0}({\bf r})
 \Biggr( \frac{1}{{\bf q}^2}\Biggr)  J^N_{0}({\bf q},{\bf p}),
 \label{eq:Mlong}
\end{eqnarray}
where Eq.~(\ref{eq:long_ME}) has been used to include the
components of ${\bf j}_e$ and ${\bf J}_N$ that are parallel to
${\bf q}$.
The longitudinal response function is obtained by dividing the cross
section integrated over the angles of the knocked-out nucleon by
the Mott cross section,
\begin{equation}
R_L =\frac{{\bf Q}^4}{\sigma_{Mott}Q^4} \int d\Omega_p \frac{4
\alpha^2 }{(2\pi)^5 }\overline{\left|{\cal M}_L \right|^2} E_f^2
p E_p,
\end{equation}
where ${\cal M}_L$ is the longitudinal amplitude of
Eq.~(\ref{eq:Mlong}).
The full calculation thus involves
a six-dimensional integration in order to obtain the amplitude
${\cal M}_L$.  Two more integrations over the angles of the
knocked-out nucleon are required in order to obtain the
response function. Results based on
the eight-dimensional integration are called ``full calculations''
in the following sections.

When the Klein-Gordon equation is used for
the electron's Coulomb distorted waves, the time component of the
current is $j^0= [E_i + E_f - 2 V(r)]/\sqrt{4E_iE_f}$
when Coulomb effects are included
and $j^0 = (E_i + E_f)/\sqrt{4E_iE_f}$ in the PWIA.
The matrix element in the Klein-Gordon case is
\begin{eqnarray}
&&{\cal M}_L^{KG}
= \int d^3r \int \frac{d^3q}{(2\pi)^2} e^{i ({\bf
Q}-{\bf q})\cdot {\bf r}} e^{i\chi({\bf r})}
f^{KG(-)}_f({\bf r}) \times  \nonumber \\
&& f^{KG(+)}_i({\bf r})
\frac{[E_i + E_f - 2 V(r)]}{\sqrt{4E_iE_f}}
 \Biggr( \frac{1}{{\bf q}^2}\Biggr)  J^N_{0}({\bf q},{\bf p}),
 \label{eq:Mlong_KG}
\end{eqnarray}
An analog of the Mott cross section based on the longitudinal current is
\begin{equation}
\sigma^{KG} = \frac{\alpha^2 E_f (E_i+E_f)^2}{E_i{\bf Q}^4}
\end{equation}
and the response function is
\begin{equation}
R_L =\frac{1}{\sigma^{KG}} \int d\Omega_p \frac{4
\alpha^2 }{(2\pi)^5 }\overline{\left|{\cal M}_L^{KG} \right|^2} E_f^2
p E_p.
\end{equation}

The longitudinal respsonse function $R_L$ is calculated using a
very simple model of the nuclear current as follows,
\begin{equation}
J_N^{\mu}({\bf q},{\bf p}) = \Biggr( \frac{p_i^{\mu} +
p_f^{\mu}}{\sqrt{4E_p(E_p-\omega)} }\Biggr) \psi({\bf q}-{\bf
p}), \label{eq:nuclearcurr}
\end{equation}
where $\psi({\bf k})$ is a gaussian wave function for a bound
nucleon,
\begin{eqnarray}
\hat{\psi}({\bf k}) = (2 \pi \beta^2)^{3/4} e^{-\beta^2 k^2/4},
\end{eqnarray}
normalized such that $\int d^3k |\psi({\bf k})|^2/(2\pi)^3 = 1$.
This simple model is used because the Coulomb corrections should
depend mainly on the electron wave functions.  Calculations are
based on the value $\beta$ = 2 fermi.

The nuclear current given above is based upon initial and final
momenta,
\begin{eqnarray}
p_f^{\mu} = \big( E_p, ~{\bf p} \big), \nonumber \\
p_i^{\mu} = \big( E_p - \omega, ~{\bf p}-{\bf q} \big),
\end{eqnarray}
where $\omega$ and ${\bf q}$ are the photon's energy and momentum.
Because of energy conservation, $E_p = M + \omega - B$, where $B
\approx .008 GeV$ is a typical binding energy of a nucleon.
Gauge-invariance must hold so Eq.~(\ref{eq:currcons}) is used to
eliminate the component of the nuclear current that is parallel
to the photon's momentum.

In the plane-wave impulse approximation, the longitudinal
response function is
\begin{eqnarray}
R_L^{PWIA} &=& \frac{pE_p}{(2\pi)^5} \int d\Omega_p \biggr|
\frac{{\bf Q} ^2{\cal M}_L}{cos{\theta_e\over 2}}\biggr|^2
\end{eqnarray}
where the longitudinal amplitude is
\begin{equation}
{\cal M}_L^{PWIA} = \delta_{\lambda_f \lambda_i}
\frac{h_{PWIA}^{0} J_{N}^0({\bf Q},{\bf p})}{{\bf Q}^2}.
\end{equation}
Using the current and wave function described above leads to
\begin{eqnarray}
R_L^{PWIA}&=& \frac{p}{(2 \pi)^3}\frac{(2E_p-\omega)^2} {4(E_p-\omega)}
\int d\Omega_p |\hat{\psi}({\bf Q} - {\bf
p})|^2 .\nonumber \\
\label{eq:R_L_PWIA}
\end{eqnarray}
 The angular integrations are straightforward, yielding
\begin{eqnarray}
R_L^{PWIA}(\omega,{\bf Q})  =
 \frac{1}{\sqrt{2\pi}} \frac{
(2E_p-\omega)^2}{4(E_p-\omega)} \frac{\beta}{|{\bf Q}|}\times
\nonumber \\
\biggr( e^{-\beta^2( |{\bf Q}| - p)^2 /2} - e^{-\beta^2( |{\bf
Q}| + p)^2 /2}\biggr).
\end{eqnarray}
Here the $PWIA$ response function is normalized so that at fixed
${\bf Q}$,  $ \int d\omega R_L({\bf Q}, \omega)\approx 1 $.


\section{Effective momentum approximation revisited}
\label{sec:ema}

As shown by Rosenfelder~\cite{Rosenfelder80} and
Traini~\cite{Traini01}, there are significant cancellations in the
Coulomb corrections when response functions are evaluated in an
effective-momentum approximation ($ema$). This approximation
usually is based on expanding the eikonal phase in a Taylor's
series about ${\bf r} = {\bf 0}$ and keeping the first two terms
as follows,
\begin{equation}
\chi({\bf r}) \approx \chi({\bf 0}) + {\bf r}\cdot \nabla
\chi({\bf 0} + \cdots.
\end{equation}
The focusing factors are approximated by their values at ${\bf
r}= {\bf 0}$ and the helicity matrix
elements are approximated by the plane-wave values. Integration
over ${\bf r}$ then gives $\delta^{(3)}({\bf q} - {\bf Q}_{eff})$,
so the longitudinal amplitude simplifies to the PWIA form as
follows,
\begin{equation}
{\cal M}_L^{ema} = 2\pi \delta_{\lambda_f \lambda_i}h_{PWIA}^{0}
J_{N}^0({\bf Q}_{eff},{\bf p})\frac{f^{D(-)}_f({\bf 0})f^{D(+)}_i({\bf 0})
}{{\bf Q}_{eff}^2}. \label{eq:ML_ema}
\end{equation}
The effective momentum involves the gradient of the eikonal phase
shift $\chi = \chi_f^{(-)} + \chi_i^{(+)}$ at the origin. Because
of cylindrical symmetry of $\chi_i^{(+)}$ about the direction
$\hat{k}_i$, $\nabla \chi_i^{(+)}$ at the origin is nonzero only
along the direction $\hat{k}_i$, and similarly $\nabla
\chi_f^{(-)}$ at the origin is nonzero only along the direction
$\hat{k}_f$. With $v_i = v_f \approx 1$, we find the same result
as Traini,
\begin{equation}
{\bf Q}_{eff} = \hat{k}_i \Big[ k_i - \delta k \Big] - \hat{k}_f
\Big[ k_f - \delta k \Big], \label{eq:Qeff}
\end{equation}
where $\delta k = V(0)$. It is correct up to first order
in the eikonal expansion because the contribution from
the gradient of eikonal correction $\chi_1$ vanishes at
the origin.

The analyses of
Rosenfelder~\cite{Rosenfelder80} and
Traini~\cite{Traini01} are based on the
approximate focusing factors, $f^{D(+)}_i \approx 1 - V(0)/E_i$ and
$f_f^{D(-)} \approx 1 - V(0)/E_f$. Coulomb effects
in the focusing factors and the effective photon propagator
cancel if one considers the
photon propagator of the transverse amplitude, which is $1/[{\bf
Q}_{eff}^2 - \omega^2]=1/[ 4 [k_i - V(0)][k_f - V(0)]sin^2{1\over
2}\theta_e]$, i.e.,
\begin{equation}
\frac{ f^{D(-)}_f({\bf 0}) f^{D(+)}_i({\bf 0})}{{\bf Q}_{eff}^2 - \omega^2}
= \frac {1}{Q^2}, \label{eq:cancel}
\end{equation}
which is the same as in the plane-wave case,
Eq.~(\ref{eq:M_PWIA}). However, the requirements of gauge
invariance that have been incorporated into the longitudinal
matrix element of Eq.~(\ref{eq:Mlong}) show that the effective
photon propagator is $1/{\bf Q}_{eff}^2$.  In that case the
cancellations still are significant but not perfect.  We find
\begin{equation}
\frac{ f^{D(-)}_f({\bf 0}) f^{D(+)}_i({\bf 0})}{{\bf Q}_{eff }^2} =
\frac{1}{{\bf Q}^2 +  \Delta},
\label{eq:part-cancel}
\end{equation}
where $\Delta$ represents a Coulomb correction to the plane-wave
result,
\begin{equation}
 \Delta = \omega^2 V(0)\frac{ [k_i + k_f -V(0)]}{[k_i - V(0)][k_f - V(0)]}
\end{equation}
As an example, for 500 MeV electrons scattering from a 25 MeV
Coulomb potential with $\theta_e$=60$^{o}$, $\Delta/{\bf Q}^2
\approx .01$ at $\omega = .125$, which is close to the
quasi-elastic peak.

 Numerical calculations based on partial-wave expansions of
Dirac-Coulomb waves also indicate that the Coulomb
corrections do not cancel to the extent that
Eq.~(\ref{eq:cancel}) would suggest.\cite{Kim05} For the
$e^+$ and $e-$ response functions the {\it ema}
analysis suggests that the same response function should be
obtained if the energy is shifted such that $E_i(e^+) + V(0) =
E_i(e^-) - V(0)$, where $V(0)$ is the electron-nucleus potential
at $r=0$.  This shift results in the same ${\bf Q}_{eff}$ for $e+$
and $e-$ scattering.
Using the gauge-invariant response function, the
correction $\Delta$ has opposite sign for $e^+$ scattering than
for $e^-$ scattering. It can produce a 4\% difference in the
$e^+$ and $e^-$ longitudinal response functions, whereas there would be no
difference based on Eq.~(\ref{eq:cancel}).  Numerical calculations
based on partial-wave expansions of Dirac-Coulomb waves
indicate that the Coulomb corrections do not cancel to the
extent that Eq.~(\ref{eq:cancel}) would suggest. In
their analysis based on the full DWBA, Kim {\it et al.}~\cite{Kim01}
have obtained results for the sum of
longitudinal and transverse responses, which differ by about 15\%
for 420 MeV $e+$ and 383 MeV $e-$.

\section{Numerical calculations for quasi-elastic scattering}
\label{sec:numerical}
Calculations of the longitudinal response function
are performed for three cases:
PWIA, ema and the full calculation.  We also use
distorted waves based on the Dirac equation and the
Klein-Gordan equation.
The calculations are based on a charge $Z=38$ and radius $R =$ 2 fermi
in the Coulomb potential of Eq.~(\ref{eq:Vofr}).  Eikonal phases
are evaluated through second order, i.e., $\chi = \chi_0 + \chi_1
+ \chi_2$ and $\omega = \omega_1 + \omega_2$.  However, the
expansion converges rapidly for the parameters and energies used
and results based on $\chi_0 + \chi_1$ and $\omega_1$ differ by
about 0.3\% at the quasi-elastic peak.

    In order to check the accuracy of the full calculation
of $R_L$,
which involves an eight-dimensional integration,
we have performed calculations
with $Z=0$ and compared with the analytical $PWIA$
result. Results are accurate to between 0.5\% to 2\% with the
integration points that have been used.

\begin{figure}[h]
\includegraphics[width=7cm,bb= 30 90 570 535,clip]{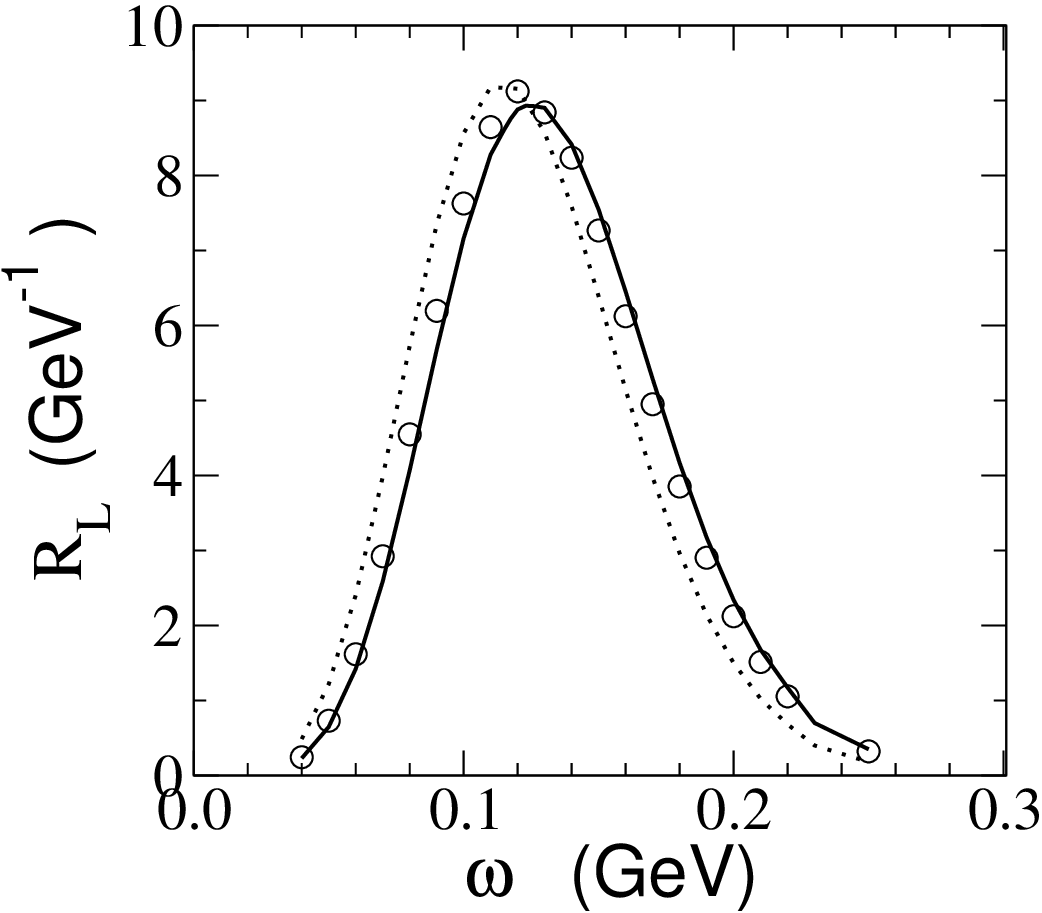}
\caption{Longitudinal response function versus the electron's
energy loss, $\omega$, calculated using Dirac distorted waves
for $e^-$ scattering at
$E= 500 MeV$ and $\theta_e =60^o$. Dotted line shows $PWIA$, solid
line shows $ema$ and circles show full calculations based on
Eq.~(\ref{eq:Mlong}).}\label{fig:RL.e-.500}
\end{figure}

\begin{figure}[h]
\includegraphics[width=7cm,bb= 52 90 578 535,clip]{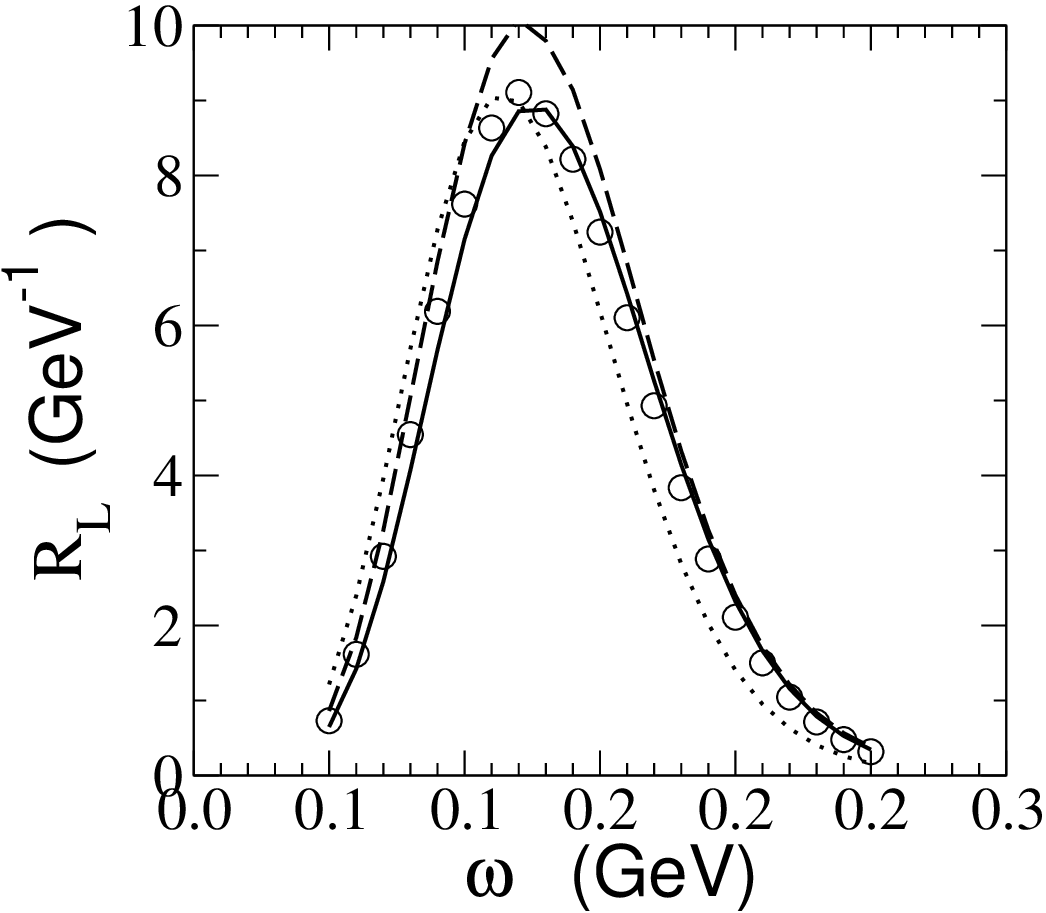}
\caption{Longitudinal response function versus the electron's energy
loss, $\omega$, calculated using Klein-Gordon
distorted waves for $e^-$ scattering at
$E= 500 MeV$ and $\theta_e =60^o$. Dotted line shows $PWIA$, solid
line shows $ema$, dashed line shows the
prescription of Ref.~\cite{Aste04a}
 and circles show full calculations based on
Eq.~(\ref{eq:Mlong_KG}).}\label{fig:RL.e-KG.500}
\end{figure}

\begin{table}[h]
\caption{Numerical results for $R_L$ for 500 MeV electrons
scattered by 60$^o$.  }
 \label{tab:RLvalues}
\begin{ruledtabular}
\begin{tabular}{cccc}
$\omega$ &  PWIA & $ema$ & Full \\
GeV   &   &   &    \\
\hline
0.040  & 0.49  &  0.23  &  0.24  \\
0.050  &  1.22 &  0.65  &  0.73  \\
0.060  &  2.41 &  1.42  &  1.62  \\
0.070  &  3.98 &  2.59  &  2.92  \\
0.080  &  5.73 &  4.07  &  4.55  \\
0.090  &  7.35 &  5.68  &  6.20  \\
0.100  &  8.56 &  7.16  &  7.63  \\
0.110  &  9.18 &  8.28  &  8.65  \\
0.120  &  9.16 &  8.87  &  9.12  \\
0.130  &  8.57 &  8.90  &  8.84  \\
0.140  &  7.59 &  8.42  &  8.24  \\
0.150  &  6.39 &  7.55  &  7.27  \\
0.160  &  5.14 &  6.45  &  6.12  \\
0.170  &  3.98 &  5.29  &  4.95  \\
0.180  &  2.96 &  4.17  &  3.85  \\
0.190  &  2.14 &  3.17  &  2.90  \\
0.200  &  1.50 &  2.34  &  2.12  \\
0.210  &  1.02 &  1.68  &  1.51  \\
0.220  &  0.68 &  1.17  &  1.06  \\
\end{tabular}
\end{ruledtabular}
\end{table}
Figure~\ref{fig:RL.e-.500} and Table~\ref{tab:RLvalues} show the longitudinal response
function for 500 MeV
electrons with scattering angle $\theta_e=60^o$
based on the use of the Dirac current and distorted waves.
In Fig.~\ref{fig:RL.e-.500} the full calculation is shown by circles,
the $ema$ calculation
based on $\delta k = V(0)$ is shown by the solid line and the
PWIA calculation is shown by the dotted line.
Figure~\ref{fig:RL.e-KG.500} shows the response
function for the same kinematics using the
Klein-Gordon current and distorted waves.  The results
are quite close to those based on the Dirac equation,
which is expected because of the similarity of eikonal wave
functions for the two cases when the spin-dependent
phases are omitted. As noted earlier, the Klein-Gordon
and Dirac results involve similar overall focusing
factors when consistent currents and wave functions
are used.
We also show by the dashed line in Fig.~\ref{fig:RL.e-KG.500}
a calculation following the prescription
of Ref~\cite{Aste04a} in which the Klein-Gordon
current is used with the Dirac focusing factors.
This produces a significantly different
result because of the extraneous factor $1 - V(0)/\bar{E}$ that
is included. The reason why the extra factor produces a large change
is because the other focusing factors largely cancel
out of the matrix element
as in Eqs.~(\ref{eq:cancel}) or (\ref{eq:part-cancel}).

\begin{figure}[h]
\includegraphics[width=7cm,bb= 52  90 558 535,clip]{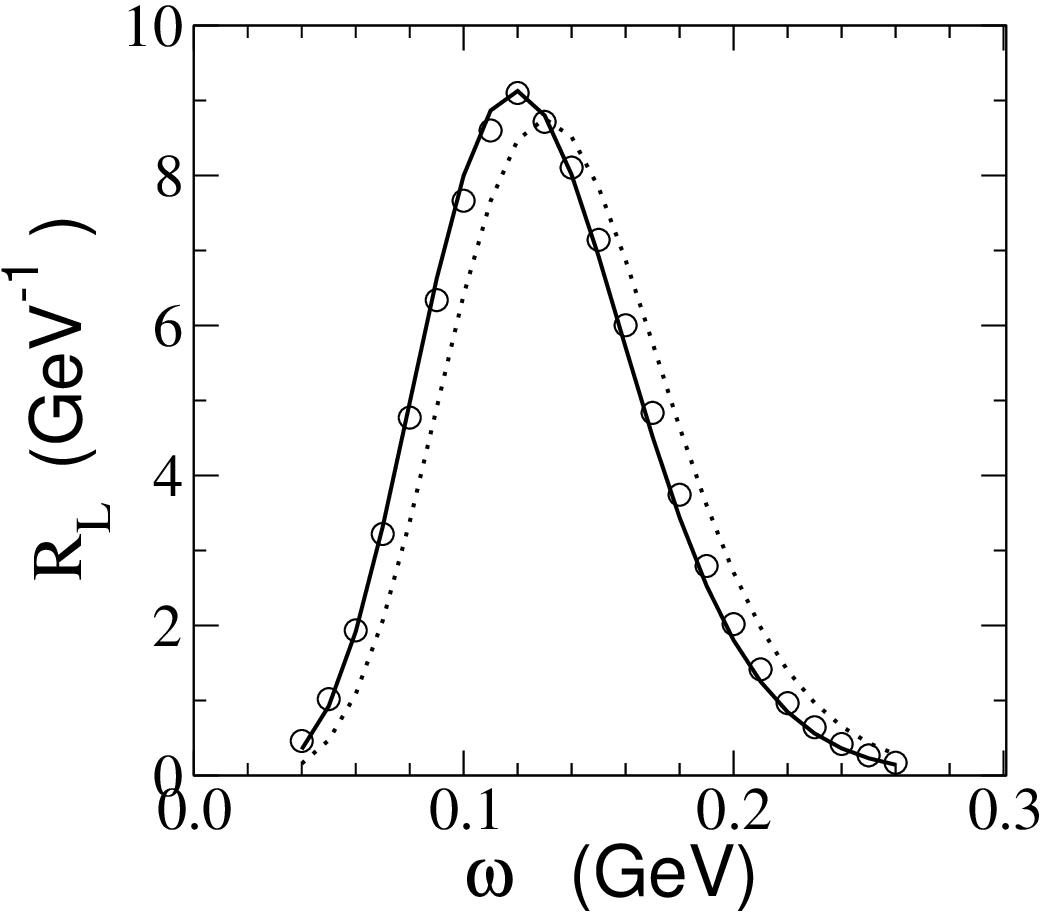}
\caption{Longitudinal response function versus the positron's
energy loss, $\omega$, for $e^+$ scattering at
$E= 540 MeV$ and $\theta_e = 60^o$.  Dotted line shows $PWIA$,
solid line shows $ema$ and circles show full calculations based
on Eq.~(\ref{eq:Mlong}).}\label{fig:RL.e+.540}
\end{figure}

  For the remainder of this section we use only the
Dirac current and Dirac wave functions.
Figure~\ref{fig:RL.e+.540} shows the
longitudinal response function for $e^{+}$
scattering at 540 MeV.  In general the $ema$
is seen in Figures~\ref{fig:RL.e-.500} and \ref{fig:RL.e+.540}
to produce a significant shift of $R_L$ away from the
PWIA result and towards the full calculation of $R_L$.
There also is good agreement between the response
functions for $e-$ and $e+$ scattering at the
energies that make ${\bf Q}_{eff}$ close to the same for both.

\begin{figure}[h]
\includegraphics[width=7cm,,bb= 52  90 578 535,clip]{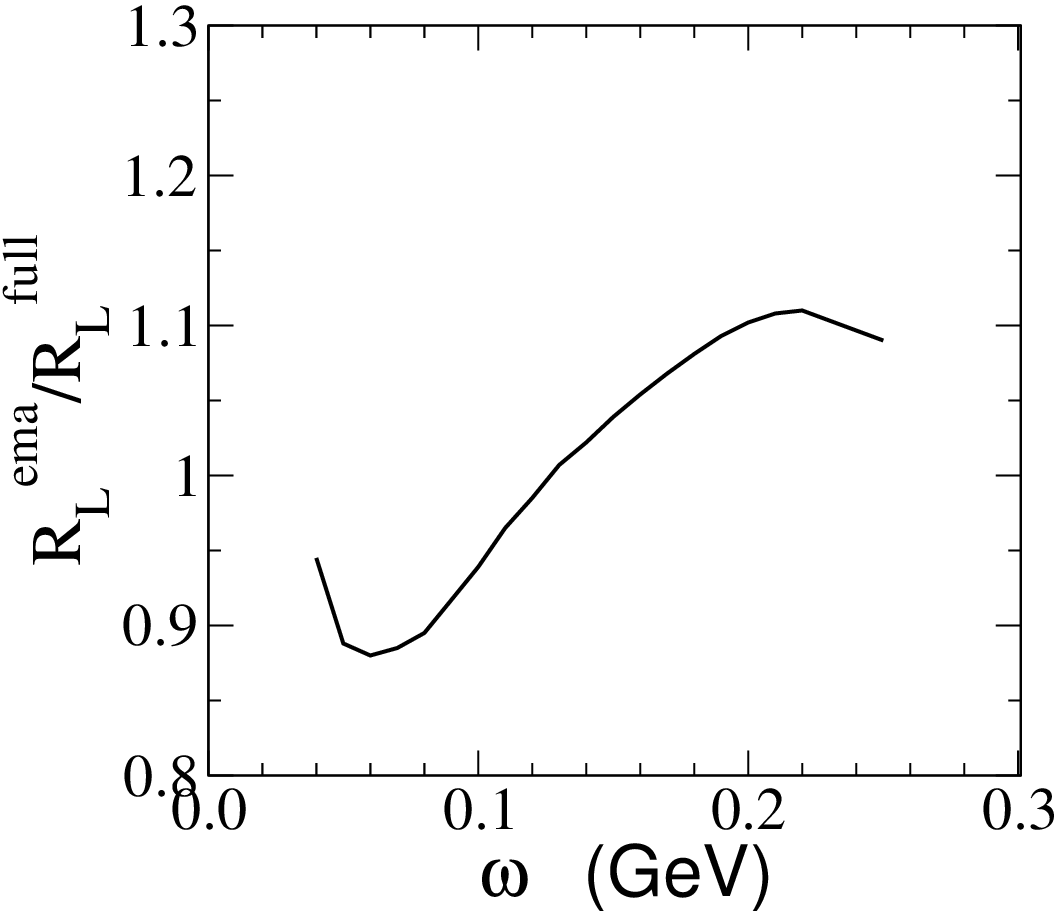}
\caption{Ratio $R_L^{ema}/R_L$ as a function of energy loss $\omega$ for 500
MeV $e^-$ scattering at angle $\theta_{e^-}=60^o$.
}\label{fig:RL_ratio.e-.500}
\end{figure}

\begin{figure}[h]
\includegraphics[width=7cm,,bb= 52  90 578 535,clip]{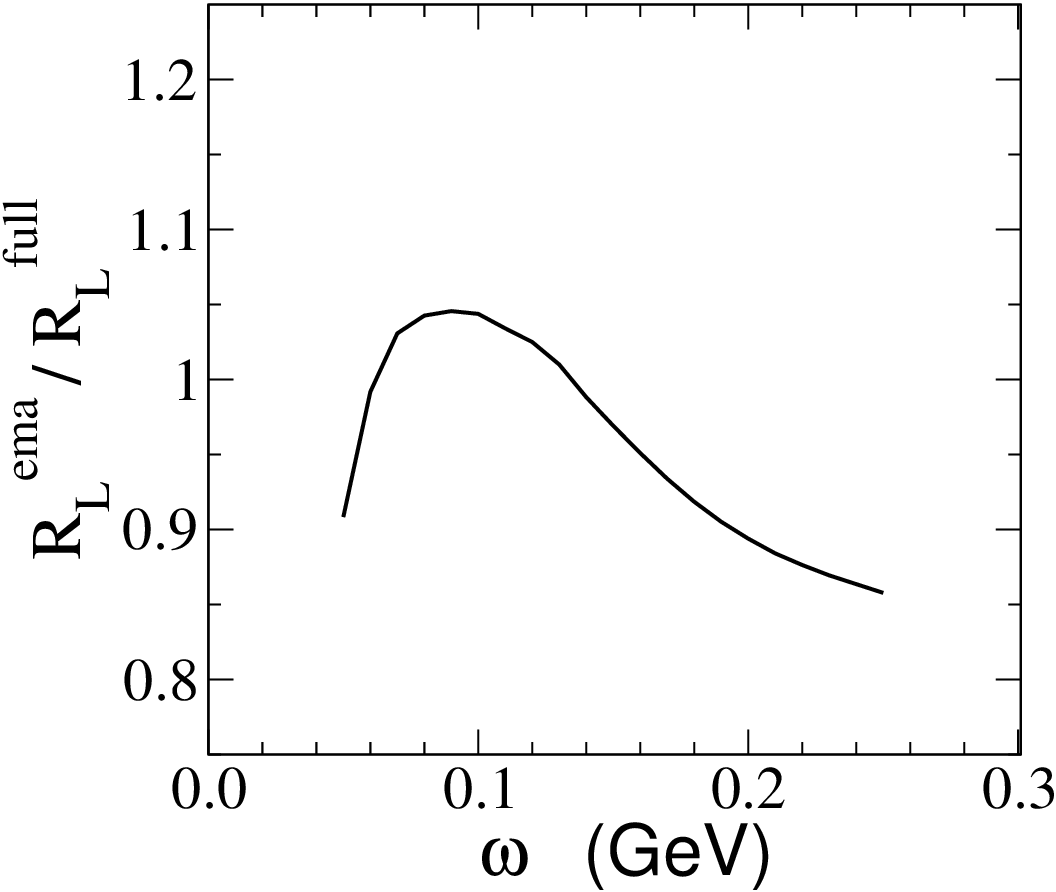}
\caption{Ratio $R_L^{ema}/R_L$ as a function of energy loss $\omega$ for 540
MeV $e^+$ scattering at angle $\theta_e=60^o$.
}\label{fig:RL_ratio.e+.540}
\end{figure}

Upon closer inspection, we find that the $ema$ result is not
precise.
Figures~\ref{fig:RL_ratio.e-.500} and \ref{fig:RL_ratio.e+.540}
show the ratios of the $ema$
response functions to the full ones for $e^-$ and $e^+$ scattering,
respectively.  Deviations of 5-10\% occur
at values of $\omega$ that are away from the quasi-elastic peak.
In the literature one finds a variety of suggestions for improving
the $ema$, such as using $\delta k = V(R)$ rather than $\delta k = V(0)$,
where $R$ is the mean nuclear radius, or using $\delta k = \overline{V}$,
where $\overline{V}$ is the average potential within a sphere of
radius $R$.  These prescriptions improve the $ema$ at some
values of $\omega$ but not all values.


%

\begin{figure}[h]
\includegraphics[width=7cm,bb= 42  90 628 535,clip]{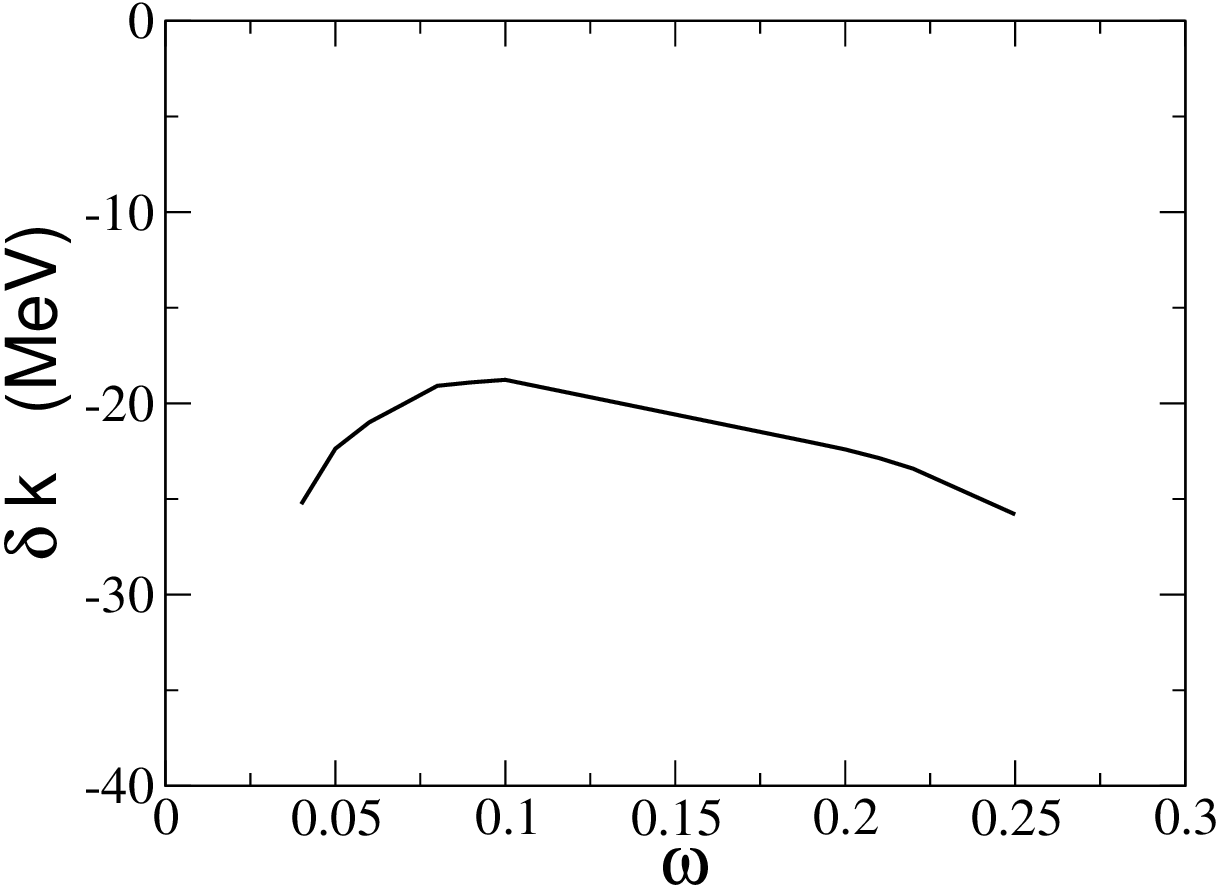}
\caption{Shift $\delta k$ used in ${\bf Q}_{eff}$ in order to
achieve the equality of Eq.~(\ref{eq:RL=RLPWIA}), as a function of
electron's energy loss, $\omega$, for 500 MeV $e^-$ scattering at
$\theta_{e^-} = 60^o$. The normalization constant is $A=0.987$.
}\label{fig:dki_fit.e-.500}
\end{figure}
\begin{figure}[h]
\includegraphics[width=7cm,bb= 42  90 628 535,clip]{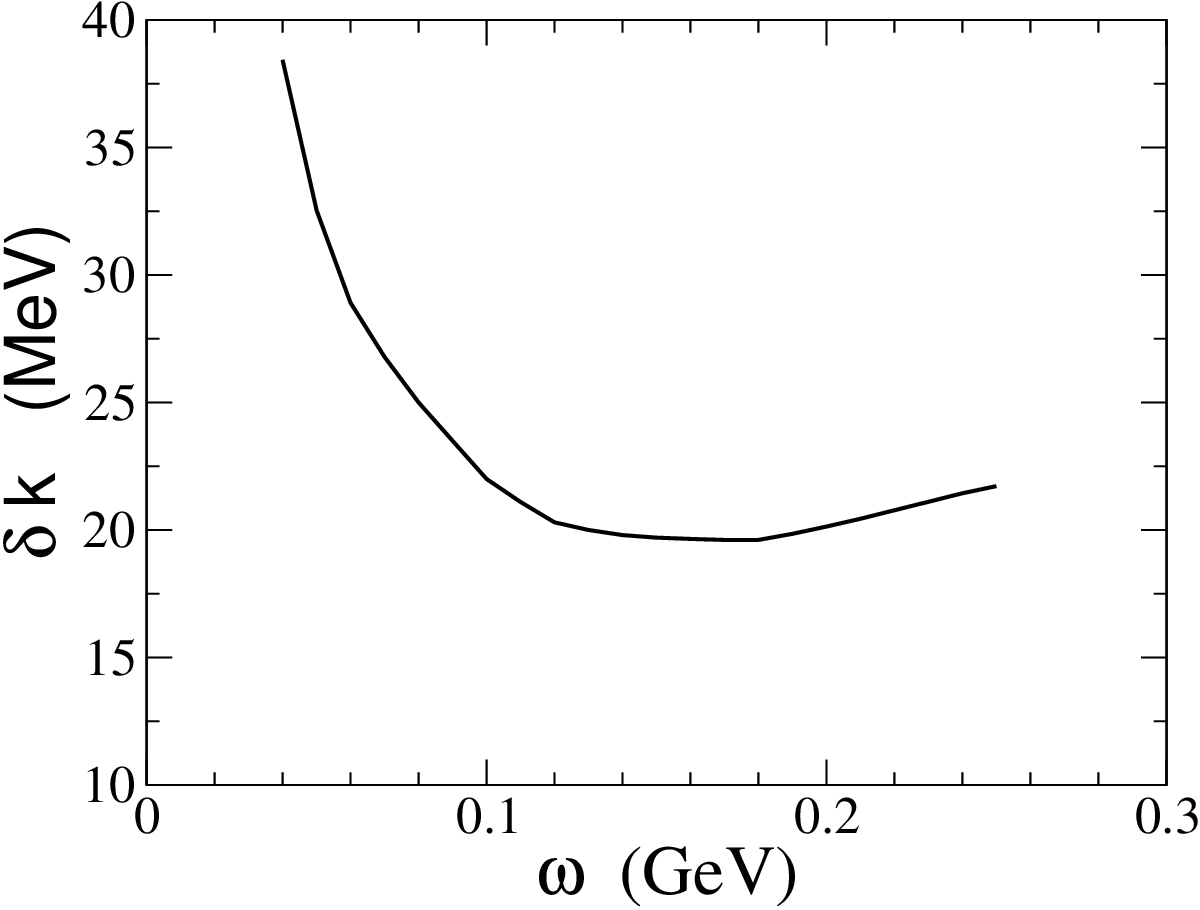}
\caption{Shift $\delta k$ used in ${\bf Q}_{eff}$ in order to
achieve the equality of Eq.~(\ref{eq:RL=RLPWIA}), as a function of
positron's energy loss, $\omega$, for 540 MeV $e^+$ scattering at
$\theta_{e^+} = 60^o$. The normalization constant is $A=0.973$.
}\label{fig:dki_fit.e+.540}
\end{figure}

A precise form of the effective-momentum approximation would be
useful for removing Coulomb corrections from experimental data in
a straightforward manner. It would allow a determination of the
PWIA response function.  In order to have a precise result,
one should determine appropriate values of the
momentum-shift function $\delta k(k_i,\omega,\theta_e)$ from which
the appropriate ${\bf Q}_{eff}$ may be calculated as in
Eq.~(\ref{eq:Qeff}). In order to determine this function, we have
fit the full response function as follows,
\begin{equation}
R_L({\bf Q},\omega)= A ~ R_L^{PWIA}({\bf Q}_{eff},\omega)
\label{eq:RL=RLPWIA}
\end{equation}
at fixed values of $k_i$ and $\theta_e$ by varying $\delta k$ and
the normalization constant $A$. The role of the normalization
constant $A$ is to ensure that $\delta k(\omega)$ is a smooth
function.  Without this parameter, $\delta k$ can take anomalous
values near the quasi-elastic peak. Figures
\ref{fig:dki_fit.e-.500} and \ref{fig:dki_fit.e+.540} show the
values of $\delta k$ that are found to yield
Eq.~(\ref{eq:RL=RLPWIA}) for scattering of $e^-$ (with A$_{e^-}$ =
0.987) and $e^+$ (with A$_{e^+}$ = 0.973). The sign of $\delta k$
is in accord with the potential $V(0)$, which is -27 MeV for
electrons and +27 MeV for positrons. However, a constant
value of $\delta k$ does not suffice.  The required values vary
with $\omega$ at a fixed value of $\theta_e=60^o$, within a range
$0.7|V(0)| < |\delta k| < 1.5|V(0)|$.
Note that the shift $\delta k$ required for
the $e^+$ response function is not simply obtained by reversing
the sign of the shift required for the $e^-$ response function.
Nevertheless, the closeness of $A$ to unity and $\delta k$ to $V(0)$
suggests that an $ema$ analysis using the fit values of $\delta
k$ would be well-motivated on physical grounds.

   If the momentum-shift function and normalization constant $A$
   are determined theoretically for a given nucleus
   based on a sophisticated model of the
nuclear current, and $R_L$ values are
available based upon experimental data, our results suggest that one
may use Eq.~(\ref{eq:RL=RLPWIA})
   with the experimentally determined $R_L$
in order to extract
   information about the undistorted response function,
$R_L^{PWIA}$.  Of course, the accuracy would depend upon
the accuracy of the L/T separation.

\section{Summary}
\label{sec:summary}

In this paper, we develop the eikonal expansion for relativistic
wave functions based on the Klein-Gordon equation and the Dirac
equation, each with a Coulombic potential.  The purpose is to
obtain some insight into the Coulomb corrections
in quasi-elastic electron scattering without a significant
loss of accuracy.
The eikonal expansion is carried out to obtain corrections up to
order $1/k^2$ to the eikonal approximation.  We show that focusing
factors are obtained in a systematic manner by use of the eikonal
expansion. Although focusing factors take somewhat different forms
for the Klein-Gordon and Dirac wave functions, equivalent results
are obtained for the current matrix elements for the two cases
because of the Coulomb correction to the Klein-Gordon current.

Based on a simple form of the Coulomb potential, analytical
results are given for the eikonal phases and thus for the eikonal
wave functions.  For scattering of electrons with energies of a
few hundred MeV or more, the eikonal expansion converges rapidly.

 Coulomb corrections in quasi-elastic electron scattering are
 considered using the analytical wave functions based on the eikonal expansion.
For the longitudinal response function, we show that the
approximate evaluation of the matrix element using the
effective-momentum approximation is modified by the requirements
of gauge invariance.  That modification causes a small
but significant difference between the $e^-$ and $e^+$ response
functions.  Moreover, the $ema$ is not sufficiently
accurate to allow a precise analysis of data because the
effective momenta $k_i-\delta k$ and $k_f-\delta k$ are not
precise when $\delta k$ is taken to be independent of $\omega$,
as it is in the usual form of $ema$. Using a simple model of
the nuclear current, we find that use of a
function $\delta k(\omega)$ can yield a precise form of the $ema$.
The analysis should be repeated for more sophisticated
models of the nuclear current with the goal
of determining the function $\delta k (\omega)$
for different models. Based on our results, we suggest that
if one is able to extract $R_L(\omega,E,\theta_e)$ from
experimental data at fixed electron beam energy, $E$, and
fixed electron scattering angle, $\theta_e$, then it may
be equated to a constant $A \approx 1$
times the PWIA response function evaluated at an effective
momentum transfer.

Whether one may extract the longitudinal response with
reasonable accuracy has not been resolved in this work.
We have addressed the transverse current matrix elements and have shown how the
spin-dependent Coulomb corrections modify the plane-wave helicity matrix
elements. There is a potentially interesting effect in the manner
that Coulomb corrections enter as a shift of
the electron scattering angle in helicity matrix elements.
However, these spin-dependent Coulomb corrections have
not been included in our calculations.  Evaluating their effects
may help to provide insight into the accuracy with which
longitudinal and transverse response functions can be separated.
Because we have only calculated the response functions based on
the longitudinal-current matrix elements, our results cannot be
compared with experimental data, or with a full DWBA analysis,
because both include an inseparable sum of longitudinal and
transverse responses.
In particular, we note that the effects of Coulomb corrections
in the helicity matrix elements of Eq.~(\ref{eq:helicity-ME})
would go in different directions for $e+$ and $e-$ scattering.

\acknowledgements
 This work was supported by the
U.S. Dept. of Energy under contracts DE-AC05-84ER40150 and
DE-FG02-93ER-40762.



\appendix
 \section{Analytical phases }
 \label{app:analytical}
 Analytical results are given for the eikonal phase shifts based on the
 potential of Eq.~(\ref{eq:Vofr}). Terms of order $m^2/E^2$ are omitted.
 In order to give the analytical results in a reasonably short
 form, we define the following quantities: $\eta = Z\alpha/v$,
 $u = \sqrt{r^2 + R^2}$, $w = \sqrt{b^2 +
R^2}$ and $\beta = tan^{-1}(w/z)$, where $\beta$ varies from $0$
to $\pi$ as $z$ varies from $-\infty$ to $\infty$. In
$\chi^{(+)}_0$, a term $\eta ln(2 \Lambda)$ is omitted, where the
eikonal integral is calculated over the range $\{-\Lambda, z\}$.
This contribution provides an irrelevant phase factor that occurs
for potentials that behave as $1/r$ as $r \rightarrow \infty$.

  The analytical results are
\begin{eqnarray}
\chi^{(+)}_0 = \eta ln\left( \frac{z + u }{w^2} \right)
 \label{eq:chi0_anal}
 \end{eqnarray}
 \begin{eqnarray}
\chi^{(+)}_1 = -\frac{\eta^2 b^2}{k w^4} \left( z + u + {1
\over 2}w \beta  \right)
\end{eqnarray}
\begin{eqnarray}
&~&\chi_2^{(+)} =-\frac{\eta^3 b^2(z+u)}{k^2w^6} \Biggr[
\frac{R^2}{2u} + \frac{(R^2 - b^2)(z+u)}{w^2}  \nonumber \\
&~&~~~~~~+ \frac{(2R^2 - b^2)\beta}{2w} \Biggr] + \frac{\eta
R^2}{2 k^2 w^4} \Bigr( \frac{z+u}{u} + \frac{w^2z}{2u^3}\Bigr)
\nonumber \\
&~& - \frac{2 \eta R^2}{ k^2 w^6} \Biggr[ \Bigr( 1 -
\frac{3b^2}{w^2}\Bigr) \Bigr( z(z+u) + \frac{w^2}{4u}(u-z)\Bigr) +
\nonumber \\
&~& \frac{b^2(z+u)}{4u} + \frac{b^2w^2z}{8u^3} \Biggr]  +
\frac{1}{4EE_2}\nabla^2\chi_0^{(+)},
 \end{eqnarray}
 \begin{eqnarray}
&~&\omega_1^{(+)} = -\frac{\eta R^2}{k w^4} \left[ z+u
-\frac{w^2}{2u} \right]
\end{eqnarray}
\begin{eqnarray}
 &~&\omega_2^{(+)} = - \frac{\eta^2}{k^2w^4}
\Biggr\{ \frac{b^2}{ 4}\left(1 + \frac{z}{u}\right)^2 + \left( 1 -
\frac{6b^2 R^2}{w^4}\right) \times \nonumber  \\
&~& \Biggr[ z(z+u) + w^2ln[2u(z+u)/w^2] +zw\beta - \frac{w^2}{2}
\Biggr]\nonumber \\ &~&
 + b^2\Bigr( 1 - \frac{2b^2}{w^2}\Bigr) \times  \Biggr[
2ln[2u(z+u)/w^2] -1
 +\frac{z\beta}{w} \Biggr]
 \nonumber \\
&~& + b^2\Biggr[ ln[2u(z+u)/w^2]+ \frac{b^2}{2u(z+u)} +
 \frac{z \beta}{2w}\left( 1 - \frac{b^2}{2w^2}\right)
 \nonumber \\
 &~&~~~~- \frac{1}{2}
 - \frac{3b^2}{4w^2}
 + \frac{b^2}{4 u^2} \Biggr] \Biggr\} \nonumber \\
 &~& +\frac{\eta^2R^2}{k^2w^4}\Biggr\{\Bigr( 1 -
 \frac{2b^2}{w^2}\Bigr) ln[2 u (u+z)/w^2]
 -\frac{w^2}{4u^2} \nonumber \\
 &~& +\frac{4b^2}{w^4} \Bigr[ z(u+z) +
 \frac{w^2}{2}\Bigr] - \frac{b^2(u+z)}{2uw^2} + \frac{b^2}{4u^2}
 \Biggr\}
\end{eqnarray}
\begin{eqnarray}
 &~&\gamma_1^{(+)} = -\frac{\eta b}{2E_2 w^2}
 \left(\frac{z+u}{u}\right)
 \end{eqnarray}
 \begin{eqnarray}
&~&\gamma_2^{(+)} = -\frac{\eta^2 b}{2k E_2 w^4} \Biggr[ 2\Bigr( 1
- \frac{2b^2}{w^2}\Bigr) \Bigr( z+u + \frac{w\beta}{2}\Bigr)
\nonumber \\ &~&~~~~~~~~~~~ + \frac{b^2}{u} + \frac{b^2\beta}{2w}
+ \frac{b^2z}{2u^2} \Biggr]
 \end{eqnarray}
 \begin{eqnarray}
&~& \delta_2^{(+)} = \frac{2 \eta R^2 b}{k E_2 w^6} \Bigr[ z + u
-\frac{w^2}{2u} - \frac{w^4}{8u^3} \Bigr]
\end{eqnarray}
 The quantity $\nabla^2 \chi^{(+)}_0$ that occurs in
$\chi_2^{(+)}$ for the Dirac case is given by
\begin{eqnarray}
\nabla^2 \chi_0^{(+)} = \frac{\eta R^2}{u^3} \Biggr[
\frac{z+2u}{(z+u)^2} - \frac{4u^3}{w^4} \Biggr]
\label{eq:gamma2_anal}
\end{eqnarray}
That term should be omitted for the Klein-Gordon case.

 Corresponding results for the incoming wave boundary condition
are obtained from Eq.~(\ref{eq:chi(-z)}), which holds for any of
the quantities in Eqs.~(\ref{eq:chi0_anal}) to
(\ref{eq:gamma2_anal}).

 \section{ Eikonal expansion for Dirac wave function}
 \label{app:dirac_eikonal_exp}

For $z \rightarrow -\infty$, the incoming wave should reduce to a
plane wave, which is realized by the boundary conditions
$\bar{\chi}^{(+)}, \bar{\gamma}^{(+)} \rightarrow 0$ as $z \rightarrow -\infty$.
Inserting Eq.~(\ref{eq:chi+}) into Eq.~(\ref{eq:uofrDirac}) leads to
\begin{eqnarray}
&~& \Biggr(E_1 - V - \frac{ (k \hat{z} + \nabla
\bar{\chi}^{(+)})^2}{E_2 - V} \nonumber \\ &~&- \sigma \cdot {\bf
p} \frac{1}{E_2-V}\sigma \cdot (k \hat{z} + \nabla
\bar{\chi}^{(+)} )
\nonumber \\
&~& - \sigma \cdot (k \hat{z} + \nabla \bar{\chi}^{(+)} )
\frac{1}{E_2-V} \sigma\cdot {\bf p} \nonumber \\ &~& - \sigma\cdot
{\bf p} \frac{1}{E_2-V} \sigma\cdot {\bf p} \Biggr) (1 -
V/E_2)^{1/2}e^{i \sigma_e \bar{\gamma}^{(+)}} = 0 .~~~
\end{eqnarray}
Momentum operators in this expression act on all quantities to
their right. Performing the indicated differentiations of the
factor $(1 - V/E_2)^{1/2}$ and multiplying by $E_2^{1/2} (E_2 -
V)^{1/2}/(2 k)$ leads to
\begin{eqnarray}
&~& \Biggr( - \frac{\partial \bar{\chi}^{(+)}}{\partial z}  -
\frac{V_{c}}{v} - \frac{(\nabla \bar{\chi}^{(+)} )^2}{2 k} + i
\frac{\nabla^2 \bar{\chi}^{(+)}}{2 k} - \frac{\sigma _e
V^{(+)}_s}{2 k}  \nonumber \\ &~&+ i \frac{\partial}{\partial z}
 + i \frac{\nabla
\bar{\chi}^{(+)}}{k} \cdot \nabla - \frac{\nabla V}{2k(E_2 -
V)}\cdot \nabla \nonumber \\ &~& + \frac{\sigma \cdot \nabla
V}{2k(E_2 - V)} \sigma \cdot \nabla + \frac{\nabla^2}{2k} \Biggr)
e^{i \sigma_e \bar{\gamma}^{(+)}} = 0,~~ \label{eq:eikonal}
\end{eqnarray}
where $v = k/E \approx 1$ and we have defined central and
spin-orbit potentials as follows,
\begin{equation}
V_{c} (r) = V(r) - \frac{V^2(r)}{2 E} + \frac{\nabla^2
V(r)}{4E(E_2 - V(r))} + \frac{3 (\nabla V(r))^2}{8 E(E_2 -
V(r))^2}, \label{eq:Vcofr}
\end{equation}
and
\begin{equation}
V^{(+)}_s(b,z) = \frac{\partial V(r)}{\partial b} - \frac{1}{E_2
-V} \frac{\partial V(r)}{\partial z} \frac{\partial
\bar{\chi}^{(+)}({\bf r}) }{\partial b}.
\end{equation}
The spin-orbit term in Eq.~(\ref{eq:eikonal}) involving potential
$V^{(+)}_s$ gives rise to the spin-dependent eikonal phase
$\bar{\gamma}^{(+)}$.

Some care is required in evaluating the derivatives of the
spin-dependent phase factor, for example, $\nabla \sigma_e =
-(\sigma_b /b)\hat{e}$ and $\nabla^2 \sigma_e = - \sigma_e/b^2$.
We find
\begin{eqnarray}
&~&- \frac{\partial \bar{\chi}^{(+)}}{\partial z}  -
\frac{V_{c}}{v} - \frac{(\nabla \bar{\chi}^{(+)} )^2}{2 k} + i
\frac{\nabla^2 \bar{\chi}^{(+)}}{2 k} -\frac{\sigma _e
V^{(+)}_s}{2 k} \nonumber \\ &~&- \frac{\partial
\bar{\gamma}^{(+)}}{\partial z} \sigma_e - \frac{\nabla
\bar{\chi}^{(+)} \cdot \nabla \bar{\gamma}^{(+)} }{k}\sigma_e
 - i \frac{\nabla V \cdot \nabla \bar{\gamma}^{(+)}}{2k(E_2 -
V)}\sigma_e \nonumber \\ &~&+ X_{3}   + i \frac{\nabla^2
\bar{\gamma}^{(+)}} {2k} \sigma_e -\frac{(\nabla
\bar{\gamma}^{(+)})^2}{2k}  \nonumber \\ &-& \frac{i}{2 k b^2}
sin(\bar{\gamma}^{(+)}) \sigma_e e^{- i \sigma_e
\bar{\gamma}^{(+)}} = 0  , \label{eq:chi_gamma}
\end{eqnarray}
where an overall factor $e^{ i \sigma_e \bar{\gamma}^{(+)}}$ is omitted, and
\begin{eqnarray}
&~&X_3= e^{-i \bar{\gamma}^{(+)} \sigma_e}\frac{\sigma\cdot \nabla
V}{2 k (E_2-V)} \sigma \cdot \nabla e^{i \bar{\gamma}^{(+)}
\sigma_e}
%
%
\end{eqnarray}
The quantity $X_3$ is found to be of order $1/k^3$ or smaller so
will be dropped in the following.

The part of Eq.~(\ref{eq:chi_gamma}) that involves $\sigma_e$ and
the part that does not involve $\sigma_e$ must vanish separately.
Using trace techniques to project out these parts produces
equations for the phases $\bar{\chi}^{(+)}$ and
$\bar{\gamma}^{(+)}$ as follows,
\begin{eqnarray}
\frac{\partial \bar{\chi}^{(+)}}{\partial z}  &=& -
\frac{V_{c}}{v} - \frac{(\nabla \bar{\chi}^{(+)} )^2}{2 k} + i
\frac{\nabla^2 \bar{\chi}^{(+)}}{2 k} - \frac{(\nabla
\bar{\gamma}^{(+)})^2}{2k}\nonumber \\
&-&  \frac{1}{2 k b^2}\sin^2(\bar{\gamma}^{(+)}),
\label{eq:dchidz}
\end{eqnarray}
\begin{eqnarray}
 \frac{\partial \bar{\gamma}^{(+)}}{\partial z}  &=& -
\frac{V^{(+)}_{s}}{2k} - \frac{\nabla \bar{\chi}^{(+)}\cdot \nabla
\bar{\gamma}^{(+)} }{ k} + i \frac{\nabla^2 \bar{\gamma}^{(+)}}{2
k} \nonumber \\ &-& \frac{i}{4 k b^2}\sin(2\bar{\gamma}^{(+)}) -
i \frac{\nabla V \cdot \nabla \bar{\gamma}^{(+)}}{2k(E_2 - V)}.
\label{eq:dgammadz}
\end{eqnarray}
One may see from the second of these equations that
$\bar{\gamma}^{(+)}$ is of order $1/k$ or smaller.  Thus, the
last two terms in Eq.~(\ref{eq:dchidz}), the last term in
Eq.~(\ref{eq:dgammadz}) and $X_3$ are of order $1/k^3$ or smaller.
We omit these terms in the following with the objective of
obtaining results correct to order $1/k^2$. We also simplify
$\sin(2\bar{\gamma}^{(+)}) \approx 2 \bar{\gamma}^{(+)}$ for the
same reason.

Integration in accord with the outgoing-wave boundary conditions
produces the basic equations
 \begin{eqnarray}
 \bar{\chi}^{(+)}({\bf r}) &=& -\int_{- \infty}^z dz' \frac{V_{c}({\bf r}')}{v} -
\int_{- \infty}^z dz' \frac{(\nabla ' \bar{\chi}^{(+)}({\bf r}')
)^2}{2 k} \nonumber \\ &~&+ i \int_{- \infty}^z  dz'
\frac{\nabla^{'2}
 \bar{\chi}^{(+)}({\bf r}')}{2 k},
\label{eq:eikonal+chi}
 \end{eqnarray}
 \begin{eqnarray}
  \bar{\gamma}^{(+)}({\bf r}) &=& -\int_{- \infty}^z dz' \frac{V^{(+)}_{s}({\bf r}')}{2k}
  \nonumber \\ &-&
\int_{- \infty}^z dz' \frac{\nabla ' \bar{\chi}^{(+)}({\bf r}')
\cdot \nabla ' \bar{\gamma}^{(+)}({\bf r}') }{2 k} \nonumber \\
&+& i \int_{- \infty}^z  dz' \frac{\Big[\nabla^{'2} - 1/b^2\Big]
 \bar{\gamma}^{(+)}({\bf r}')}{2 k} ,
 \nonumber \\
\label{eq:eikonal+gamma}
 \end{eqnarray}
where ${\bf r}' = ({\bf b}, z')$.

With incoming-wave boundary conditions, the wave function must
become a plane wave as $z \rightarrow +\infty$ and is written as
\begin{eqnarray}
u^{(-)}({\bf r}) = \left( 1 - \frac{V}{E_2}\right) ^{1/2} e^{ik z
} e^{-i \bar{\chi}^{(-)}}e^{-i\sigma_e \bar{\gamma}^{(-)}}.
\end{eqnarray}
and the complex phases $\bar{\chi}^{(-)}({\bf r})$ and
$\bar{\gamma}^{(-)}({\bf r})$ obey
 \begin{eqnarray}
  \bar{\chi}^{(-)}({\bf r}) &=& -\int^{ \infty}_z  dz' \frac{V_{c}({\bf r}')}{v} -
\int^{ \infty}_z  dz' \frac{(\nabla ' \bar{\chi}^{(-)}({\bf r}')
)^2}{2 k} \nonumber \\ &-& i \int^{ \infty}_z  dz' \frac{
\nabla^{'2}
 \bar{\chi}^{(-)}({\bf r}')}{2 k}
\label{eq:eikonal-chi}
\end{eqnarray}
 \begin{eqnarray}
 \bar{\gamma}^{(-)}({\bf r}) &=& -\int^{ \infty}_z  dz' \frac{V^{(-)}_{s}({\bf r}')}{2k}
 \nonumber \\ &~&-
\int^{ \infty}_z  dz' \frac{\nabla ' \bar{\chi}^{(-)}({\bf r}')
\cdot \nabla '\bar{\gamma}^{(-)}({\bf r}') }{2 k} \nonumber \\
&~&- i \int^{ \infty}_z  dz' \frac{\Big[\nabla^{'2} - 1 /b^2 \Big]
 \bar{\gamma}^{(-)}({\bf r}')}{2 k}.
 \label{eq:eikonal-gamma}
 \end{eqnarray}
where $V_s^{(-)}$ is
\begin{equation}
V^{(-)}_s(b,z) = \frac{\partial V(r)}{\partial b} + \frac{1}{E_2
-V} \frac{\partial V(r)}{\partial z} \frac{\partial
\bar{\chi}^{(-)}({\bf r}) }{\partial b}.
\end{equation}

 The eikonal expansion is the iterative
solution appropriate to large $k$ of Eqs.~(\ref{eq:eikonal+chi})
and (\ref{eq:eikonal+gamma}) or Eqs.~(\ref{eq:eikonal-chi}) and
(\ref{eq:eikonal-gamma}). The expansion for $\bar{\chi}^{(+)}$
initially takes the form
\begin{equation}
\bar{\chi}^{(+)} = \bar{\chi}^{(+)}_0 + \bar{\chi}^{(+)}_1 +
\bar{\chi}^{(+)}_2 + \cdots,
\end{equation}
where each term in the series is smaller than the previous one.
The ``barred'' phases are in general
complex with the exception that $\bar{\chi}^{(+)}_0$ is real when the potential is real.

Lowest order terms may be obtained from Eqs.~(\ref{eq:eikonal+chi})
 and (\ref{eq:eikonal-chi}) by keeping just the leading terms on the right
side,
\begin{eqnarray}
\bar{\chi}^{(+)}_0({\bf r}) &=& -\frac{1}{v}\int_{- \infty}^{z}
dz' V_{c}(r'),\nonumber \\ \bar{\chi}^{(-)}_0({\bf r}) &=&
-\frac{1}{v}\int^{ \infty}_{z} dz'V_{c}(r').
 \label{eq:chi_0}
\end{eqnarray}
It should be noted that the same symbol $\bar{\chi}_0^{(+)}$ was
used in the Klein-Gordon case but the meaning is different here
because of the extra terms in $V_c(r)$, Eq.~(\ref{eq:Vcofr}).
Using the same iterative procedure that was used to obtain
Eqs.~(\ref{eq:KG-chi_1}) and (\ref{eq:KG-chi_2}) leads to
\begin{eqnarray}
\bar{\chi}^{(+)}_1({\bf r}) &=& - \frac{1}{2k}\int_{- \infty}^{z}
dz' (\nabla' \bar{\chi}^{(+)}_0({\bf r}'))^2 \nonumber \\ &+&
\frac{i}{2k} \int_{- \infty}^{z} dz'{\nabla '}^2
\bar{\chi}^{(+)}_0({\bf r}'), \label{eq:D-chi_1}
\end{eqnarray}
\begin{eqnarray}
\bar{\chi}^{(+)}_2({\bf r}) &=& - \frac{1}{2k} \int_{- \infty}^{z}
dz'\Bigr[ 2\nabla ' \bar{\chi}^{(+)}_0({\bf r}')\cdot \nabla'
\bar{\chi}^{(+)}_1({\bf r}') \nonumber \\ &+& (\nabla'
\bar{\chi}^{(+)}_1({\bf r}'))^2 \Bigr] + \frac{i}{2k} \int_{-
\infty}^{z} dz' {\nabla '}^2 \bar{\chi}^{(+)}_1({\bf r}').
\label{eq:D-chi_2}\nonumber \\
\end{eqnarray}

For the spin-dependent phase, $\bar{\gamma}^{(+)}$, the iterative expansion is
\begin{equation}
\bar{\gamma}^{(+)} = \bar{\gamma}_1^{(+)} + \bar{\gamma}_2^{(+)} + \cdots
\end{equation}
where
\begin{equation}
\bar{\gamma}_1^{(+)}({\bf r}) = - \frac{1}{2k} \int_{-\infty }^z
dz' V^{(+)}_s({\bf b},z'),
\end{equation}
\begin{eqnarray}
\bar{\gamma}_2^{(+)} ({\bf r}) &=& - \frac{1}{2k} \int_{-
\infty}^z dz' \nabla' \bar{\chi}^{(+)} ({\bf r}') \cdot \nabla '
\bar{\gamma}_1^{(+)} ({\bf r}')
\nonumber \\
&+& \frac{i}{2k} \int_{- \infty}^z dz' \Big(\nabla^{'
2}-\frac{1}{b^2}\Big) \bar{\gamma}_1^{(+)} ({\bf r}').
\end{eqnarray}

Complex eikonal phases $\bar{\chi}^{(+)}$ and $\bar{\gamma}^{(+)}$
are decomposed into real and imaginary parts ( the ``unbarred'' phases) as follows
\begin{eqnarray}
\bar{\chi}^{(+)}= \chi^{(+)} +  i \omega^{(+)}
\nonumber \\
\bar{\gamma}^{(+)}= \gamma^{(+)} +  i \delta^{(+)}.
\end{eqnarray}
 Because the central and spin-dependent potentials have terms that
involve various powers of $1/k$, the ``barred'' phases that are
obtained from the iterative expansion above are not ordered
systematically. Therefore a second expansion is performed to
obtain terms that are proportional to powers of $1/k $.  This
leads to Eqs.~(\ref{eq:chi_exp}) to (\ref{eq:eikonal_exp}) of the text.

\bibliography{basename of .bib file}

\end{document}